\documentclass[aps, prf,onecolumn,superscriptaddress,amsmath,amssymb]{revtex4-1}
\usepackage{graphicx}
\usepackage{epstopdf, epsfig}
\usepackage{dcolumn}   
\usepackage{bm}        
\usepackage{verbatim} 
\usepackage{tikz}
\usepackage{overpic}
\usepackage[outline]{contour}
\contourlength{1.2pt}
\usepackage{subcaption}
\usepackage{amsmath}
\newcommand{\beq}{\begin{equation}}
\newcommand{\eeq}{\end{equation}}
\newcommand{\un}[1]{\,\mathrm{#1}}
\newcommand{\lb}{\left(}
\newcommand{\rb}{\right)}

\begin{document}

\title{The effect of {inner} swirl on confined co-axial flow}
\author{GP Benham}
\affiliation{
   Mathematical Institute, University of Oxford, Andrew Wiles Building, Radcliffe Observatory Quarter, Woodstock Road, Oxford OX2 6GG United Kingdom
   }
   \author{IJ Hewitt}
\affiliation{
   Mathematical Institute, University of Oxford, Andrew Wiles Building, Radcliffe Observatory Quarter, Woodstock Road, Oxford OX2 6GG United Kingdom
   }
   \author{CP Please}
\affiliation{
   Mathematical Institute, University of Oxford, Andrew Wiles Building, Radcliffe Observatory Quarter, Woodstock Road, Oxford OX2 6GG United Kingdom
   }
   \author{P Bird}
\affiliation{
VerdErg Renewable Energy Limited, 6 Old London Rd, Kingston upon Thames KT2 6QF, United Kingdom
   }
\date{\today}

\begin{abstract}
We study the problem of mixing between core and annular flow in a pipe, examining the effect of a swirling core flow. Such flows are important across a range of applications, including jet pumps, combustion chambers and aerospace engineering. Previous studies show that swirl can increase shear layer growth rates and, in the case of confining walls, reduce flow separation. 
However, the effect of swirl on pressure loss in a confined flow is uncertain. To address this, we develop a simplified model that approximates the axial flow profile as a linear shear layer separating uniform-velocity core and annular streams. The azimuthal flow profile is approximated as a solid body rotation within the core region, and a parabolic mixing profile within the shear layer. This model shows good agreement with computational turbulence modelling, whilst its simplicity and low computational cost make it ideal for benchmark predictions and design purposes. Using this model, we confirm that a swirling core is useful for increasing shear layer growth rates, but find that it is detrimental to pressure recovery. This has important implications for the design of diffusers that incorporate swirling flows. We use the model to describe the slow recirculation region that can form along the pipe axis for sufficiently large swirl, by approximating it as a stagnant zone with zero velocity. The criteria for the development of such a region are established in terms of the pipe expansion angle and inflow velocity profile.  
 \end{abstract}

\maketitle

\section{Introduction}

There are many industrial applications where swirl is added to jet flows and annular flows to improve performance.
Most commonly, swirl is used in combustion chambers to increase fuel mixing rates and to stabilise the flame \cite{lilley1977swirl, lee2005experimental}. There have also been studies which indicate that swirl can be used to improve plasma jet cutting performance \cite{gonzalez1999theoretical}, to increase the efficiency of a jet pump \cite{guillaume2004improving}, and to reduce flow separation in diffusers \cite{fox1971effects}. In addition, swirl is present in a variety of propulsion systems, such as in jet engines and turbomachinery, and plays an important role in the interaction between aircraft wakes \cite{bilanin1977vortex}. 
Our goal in this study is to examine the effect of swirl on mixing of confined co-axial flows.

Existing experimental studies of \textit{unconfined} jets have shown that swirl (in the jet) can significantly increase entrainment rates and consequently the growth rate of the jet, both for compressible and incompressible fluids \cite{naughton1997experimental, gilchrist2005experimental, semaan2013three}. 
Besides other experimental and numerical studies of unconfined jets \cite{leschziner1984computation, gibson1986calculation, elsner1987characteristics, farokhi1989effect, grauer1991numerical, herrada2003vortex, liang2005experimental, shtern2011development, shtern2011development2, semaan2013three, oberleithner2014impact}, 
there have also been some that consider the effect of swirl on annular flows and confined flows.

In the case of annular flows, \citet{lee2005experimental} studied the effect of swirl on the characteristics of an unconfined co-axial annular flow. In their experiments, both the core flow and the annular flow had swirl components, not necessarily in the same direction. They showed that if the swirl is sufficiently strong, a stagnation point and recirculation region can develop inside the core flow. Other studies \cite{chigier1967experimental, escudier1985recirculation, champagne2000experiments, nayeri2000investigation} have discussed this recirculation region and relate it to the phenomenon of vortex breakdown. In the context of combustion chambers, the recirculation region is important because it has a stabilising effect on the flame. In this region, the flame speed and flow velocity are equalised due to the relatively low velocity. Both co-swirling and counter-swirling annular flows have been investigated experimentally \cite{ribeiro1980coaxial, durbin1996studies, gupta2001swirl, merkle2003effect, vanierschot2018double} with the general conclusion that counter-swirl is more stabilising.

In the case of confined flows, there are a number of studies which consider the effect of swirl on the performance of a diffuser. 
The experiments of \citet{fox1971effects} showed that by adding a solid body rotation to the inflow of a diffuser with flow separation, it was possible to reduce separation and increase pressure recovery significantly. By contrast, for unseparated flows, it was shown that swirl makes little difference to pressure recovery. The numerical simulations of \citet{hah1983calculation} investigated the effects of a Rankine-type rotation and a solid-body rotation at the inlet of a diffuser, finding similar results to Fox \& McDonald. Hah suggested that swirl reduces flow separation by means of the centrifugal force, which presses the boundary layer to the pipe wall. Both \citet{fox1971effects} and \citet{hah1983calculation} showed that performance decreases for large swirl intensities, and attribute this behaviour to the formation of a recirculation region when the swirl number is large. \citet{so1967vortex} studied the recirculation region inside a conical diffuser in detail and presented a simple model based on integral equations of mass, axial momentum, angular momentum, and moment of axial momentum. However, there was a significant discrepancy between the model predictions and experimental results, which was attributed to the assumption of constant viscosity. 
The experiments of \citet{nayeri2000investigation} studied a shear layer between confined swirling co-axial flows. Both co-swirling and counter-swirling flows were investigated and it was found that, in both cases, the shear layer growth rates were larger than in the absence of swirl, though the counter-rotating case had the largest growth rates overall. The formation of a recirculation region was not studied. Nor was the effect of different pipe geometries.


{In this study we present a simple model for confined co-axial flow with a swirling core, and we use the model to investigate the effect of swirl on the flow characteristics in a variety of pipe geometries.  
We study the effect of a swirling core on the pressure recovery in a diffuser, which is not addressed in the literature. 
In particular, we show that whilst a swirling core is useful for increasing shear layer growth rates and mixing the flows over a shorter length scale, it is detrimental to pressure recovery. We confirm that under certain flow conditions, a recirculation region can form along the pipe axis, and we use our model to characterise the onset and size of this region in terms of the inflow conditions and the pipe expansion angle.

The model is based on a previous model for non-swirling shear layers \cite{benham2018turbulent}, and uses a similar approach to \citet{so1967vortex}, where the governing equations are integral equations of mass, axial momentum and angular momentum.}

The model is described in Section \ref{secmod}, and compared to computational turbulence modelling in Section \ref{cfdcomp}. In Section \ref{predictions} we use the model to address the question of how swirl affects mixing and pressure loss in diffusers of different shapes, and we close with a summary in Section \ref{secsumm}.




\section{Mathematical model}\label{secmod}

In this section we describe the flow scenario we consider and present a simple model, which is an extension of the model presented by \citet{benham2018turbulent} for the case of no swirl. 
The model in that study considers an inflow composed of inner and outer streams of different speeds, with a velocity jump between them that evolves into a shear layer downstream.
The model presented here addresses the same inflow profile for an axisymmetric pipe but with the inner (core) stream rotating. This is achieved by introducing an angular velocity component which is governed by equations derived from integrating conservation of angular momentum. Furthermore, we account for radial pressure gradients induced by the swirl as a consequence of the conservation of radial momentum. The shear layer growth equation is modified to account for the additional effect of the azimuthal shear stress induced by the swirl. 

We also present an extended version of the model which addresses the development of a turbulent boundary layer near the pipe wall, and a turbulent symmetry layer near the pipe axis. 
This extended model is discussed later in this section and in the Supplemental Materials \cite{supp}.

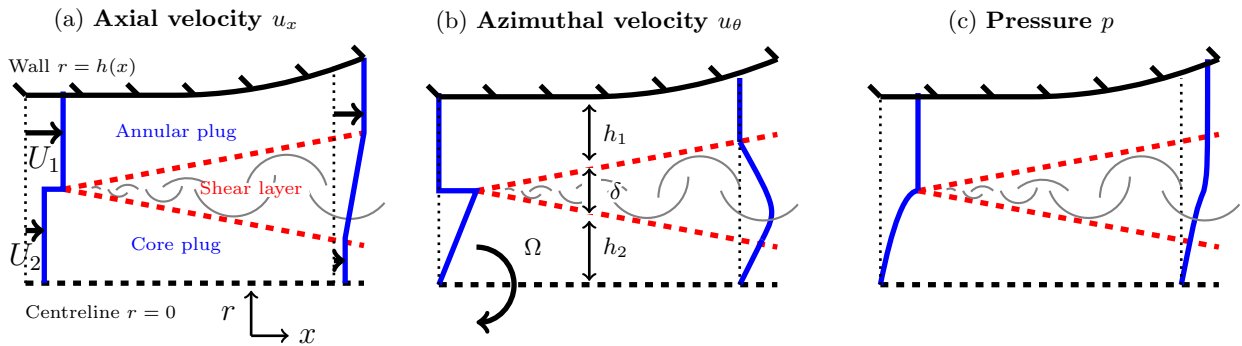
\begin{figure}
\centering
\begin{subfigure}{0.3\textwidth}
\begin{tikzpicture}[scale=0.5]
\node at (5,7) {(a) \bf Axial velocity $u_x$};
\node at (3,-0.75) {\scriptsize Centreline $r=0$};
\node at (2.25,5.75) {\scriptsize Wall $r=h(x)$};
\node at (5,4) {\scriptsize \color{blue} Annular plug};
\node at (5,1) {\scriptsize \color{blue} Core plug};
\draw [thick,gray] (3,2.5) arc [radius=0.2, start angle=45, end angle= 120];
\draw [thick,gray] (3.2,2.5) arc [radius=0.3, start angle=200, end angle= 320];
\draw [thick,gray] (4.0,2.5) arc [radius=0.4, start angle=45, end angle= 140];
\draw [thick,gray] (4.2,2.5) arc [radius=0.5, start angle=200, end angle= 350];
\draw [thick,gray] (6.0,2.5) arc [radius=1.0, start angle=45, end angle= 130];
\draw [thick,gray] (5.5,2.5) arc [radius=1.1, start angle=200, end angle= 360];
\draw [thick,gray] (9,2.6) arc [radius=1.1, start angle=15, end angle= 180];
\draw [thick,gray] (8.5,2.5) arc [radius=1.2, start angle=200, end angle= 320];
\draw[dashed,line width=2] (1,0) -- (10,0); 
\draw[red,dashed,line width=2] (2,5/2) -- (10,4); 
\draw[red,dashed,line width=2] (2,5/2) -- (10,1); 
\draw[blue, line width=2] (3/2, 0) -- (3/2,5/2) -- (2,5/2) -- (2,5.1) ; 
\draw[blue, line width=2] (9.5, 0) -- (9.5,1.2) -- (10,4) -- (10,6) ; 
\draw[->,line width=2] (1,4) -- (2,4); 
\draw[->,line width=2] (1,1.4) -- (3/2,1.4); 
\draw[->,line width=2] (9.2,1.2/2) -- (9.5,1.2/2); 
\draw[->,line width=2] (9.2,4.5) -- (10,4.5); 
\draw[dotted,line width=1] (1,0) -- (1,5); 
\draw[dotted,line width=1] (9.2,0) -- (9.2,52/9); 
\node at (1.5,3.3) {\large $U_1$};
\node at (1,0.7) {\large $U_2$};
\draw[black,line width=2,-] (1,5) parabola bend (5,5) (10,6);
\draw[line width=2] (1,5) -- (0.7,5.3); 
\draw[line width=2] (2.5,5) -- (2.2,5.3); 
\draw[line width=2] (4,5) -- (3.7,5.3); 
\draw[line width=2] (5.5,5) -- (5.2,5.3); 
\draw[line width=2] (7,5.2) -- (6.7,5.5); 
\draw[line width=2] (8.5,5.5) -- (8.2,5.8); 
\draw[line width=2] (10,6) -- (9.7,6.3); 
\draw[->,line width=1] (7,-1.4) -- (8,-1.4);
\draw[->,line width=1] (7,-1.4) --(7,-0.2);
\node at (8.5,-1.4) {\large $x$};
\node at (6.4,-0.8) {\large $r$};
\node at (3,-1) {};
\node at (7,2.5) {\scriptsize \contour{white}{\color{red} Shear layer}};
\end{tikzpicture}
\end{subfigure}
\begin{subfigure}{0.3\textwidth}
\begin{tikzpicture}[scale=0.5]
\node at (5,7) {(b) \bf Azimuthal velocity $u_\theta$ };
\draw[dashed,line width=2] (1,0) -- (10,0); 
\draw[red,dashed,line width=2] (2,5/2) -- (10,4); 
\draw[red,dashed,line width=2] (2,5/2) -- (10,1); 
\draw[blue, line width=2] (1, 0) -- (2,2.5) -- (1,2.5) -- (1,5.1) ; 
\draw[blue, line width=2] (9, 0) -- (9.6,1.2) ; 
\draw[blue, line width=2]  (9,3.8) -- (9,5.5) ; 
\draw[line width=2,blue]  (9.6,1.2) .. controls (10,2) .. (9,3.8);
\draw [line width=2, ->] (2,1) arc [radius=1, start angle=90, end angle= -90];
\node at (3.5,1) {$\Omega$};
\draw[dotted,line width=1] (1,0) -- (1,5); 
\draw[dotted,line width=1] (9,0) -- (9,5.5); 
\draw [thick,gray] (3,2.5) arc [radius=0.2, start angle=45, end angle= 120];
\draw [thick,gray] (3.2,2.5) arc [radius=0.3, start angle=200, end angle= 320];
\draw [thick,gray] (4.0,2.5) arc [radius=0.4, start angle=45, end angle= 140];
\draw [thick,gray] (4.2,2.5) arc [radius=0.5, start angle=200, end angle= 350];
\draw [thick,gray] (6.0,2.5) arc [radius=1.0, start angle=45, end angle= 130];
\draw [thick,gray] (5.5,2.5) arc [radius=1.1, start angle=200, end angle= 360];
\draw [thick,gray] (9,2.6) arc [radius=1.1, start angle=15, end angle= 180];
\draw [thick,gray] (8.5,2.5) arc [radius=1.2, start angle=200, end angle= 320];
\draw[black,line width=2,-] (1,5) parabola bend (5,5) (10,6);
\draw[line width=2] (1,5) -- (0.7,5.3); 
\draw[line width=2] (2.5,5) -- (2.2,5.3); 
\draw[line width=2] (4,5) -- (3.7,5.3); 
\draw[line width=2] (5.5,5) -- (5.2,5.3); 
\draw[line width=2] (7,5.2) -- (6.7,5.5); 
\draw[line width=2] (8.5,5.5) -- (8.2,5.8); 
\draw[line width=2] (10,6) -- (9.7,6.3); 
\fill [white] (4.8,1.9) rectangle (5.3,3.1);
\draw[line width=1, <->] (5,0.1) -- (5,1.7); 
\draw[line width=1, <->] (5,1.9) -- (5,3.1); 
\draw[line width=1, <->] (5,3.3) -- (5,4.8); 
\node at (5.7,1) {\small \contour{white}{$h_2$}};
\node at (5.7,2.5) {\small \contour{white}{$\delta$}};
\node at (5.7,4) {\small \contour{white}{$h_1$}};
\node at (3,-1.5) {};
\node at (0.5,0) {};
\end{tikzpicture}
\end{subfigure}
\begin{subfigure}{0.3\textwidth}
\begin{tikzpicture}[scale=0.5]
\node at (5,7) {(c) \bf Pressure $p$};
\draw[dashed,line width=2] (1,0) -- (10,0); 
\draw[red,dashed,line width=2] (2,5/2) -- (10,4); 
\draw[red,dashed,line width=2] (2,5/2) -- (10,1); 
\draw[blue, line width=2] (2,2.5) -- (2,5.1) ; 
\draw[blue, line width=2] (2,2.5) parabola (1, 0)  ; 
\draw[blue, line width=2] (9,0) .. controls (9.4,2) .. (9.6,2.5) .. controls (9.7,3) .. (9.7,6) ;
\draw[dotted,line width=1] (1,0) -- (1,5); 
\draw[dotted,line width=1] (9,0) -- (9,5.5); 
\draw [thick,gray] (3,2.5) arc [radius=0.2, start angle=45, end angle= 120];
\draw [thick,gray] (3.2,2.5) arc [radius=0.3, start angle=200, end angle= 320];
\draw [thick,gray] (4.0,2.5) arc [radius=0.4, start angle=45, end angle= 140];
\draw [thick,gray] (4.2,2.5) arc [radius=0.5, start angle=200, end angle= 350];
\draw [thick,gray] (6.0,2.5) arc [radius=1.0, start angle=45, end angle= 130];
\draw [thick,gray] (5.5,2.5) arc [radius=1.1, start angle=200, end angle= 360];
\draw [thick,gray] (9,2.6) arc [radius=1.1, start angle=15, end angle= 180];
\draw [thick,gray] (8.5,2.5) arc [radius=1.2, start angle=200, end angle= 320];
\draw[black,line width=2,-] (1,5) parabola bend (5,5) (10,6);
\draw[line width=2] (1,5) -- (0.7,5.3); 
\draw[line width=2] (2.5,5) -- (2.2,5.3); 
\draw[line width=2] (4,5) -- (3.7,5.3); 
\draw[line width=2] (5.5,5) -- (5.2,5.3); 
\draw[line width=2] (7,5.2) -- (6.7,5.5); 
\draw[line width=2] (8.5,5.5) -- (8.2,5.8); 
\draw[line width=2] (10,6) -- (9.7,6.3); 
\node at (3,-1.5) {};
\node at (-0.5,0) {};
\end{tikzpicture}
\end{subfigure}
\caption{Schematic diagram of our simple model for a swirling slower central flow mixing with a non-swirling faster outer flow, showing axial profiles of axial velocity $u_x$, azimuthal velocity $u_\theta$ and pressure $p$. We indicate the position of the shear layer with red dashed lines. The aspect ratios of the diagrams are exaggerated for illustration purposes.  \label{schem}}
\end{figure}

The flow scenario is shown in Figure \ref{schem}, in which we illustrate our coordinate system $(x,r)$. We consider {time-averaged} axisymmetric flow in a long thin cylindrical pipe $0<r< h(x)$, where the rate of change of the pipe radius is small. The inflow is composed of a core swirling flow, with axial velocity $U_2(0)$ and solid body rotation at angular velocity $\Omega(0)$, and an annular non-swirling flow with axial velocity $U_1(0)$. A shear layer forms between the flows and grows downstream, as illustrated by red dashed lines in Figure \ref{schem}.
Throughout this study, we restrict our attention to the case where the axial velocity of the core flow is smaller than that of the annular flow $U_2<U_1$. The opposite case is prone to asymmetric instabilities, such as the Coanda effect \cite{tritton2012physical}, which are not included in our model. The flow scenario in this study is motivated by a hydropower application, in which the core swirling flow emerges from a turbine, and it is desired to mix this flow with the annular flow with minimal pressure loss.

We approximate the axial flow profile $u_x$ by decomposing it into two uniform-velocity plug regions separated by a shear layer in which the velocity varies linearly between $U_1(x)$ and $U_2(x)$. 
Therefore, the axial velocity profile is approximated by
\beq
{
u_x(x,r)=\begin{cases}
U_{2} &: 0<r<h_2,\\
U_2 +  \varepsilon_r\lb r-h_2 \rb &: h_2 <r<h -h_1 ,\\
U_{1}  &:h -h_1 <r<h,
\end{cases}\label{piecewise}
}
\eeq
where $h_1(x)$ and $h_2(x)$ are the widths of the plug regions, $\delta=h-h_1-h_2$ is the width of the shear layer, and $\varepsilon_r=(U_1-U_2)/\delta$ is the shear rate. 

Similarly to the axial velocity, we approximate the azimuthal velocity profile by decomposing it into a core swirling region and an annular region with no swirl separated by an azimuthal shear layer in which the azimuthal velocity varies quadratically (motivated by CFD calculations which are discussed later). The approximate azimuthal velocity profile is taken to be
\beq
{
u_\theta(x,r)=\begin{cases}
\Omega  r &: 0<r<h_2,\\
(h - h_1 - r ) (\frac{h_2 \Omega}{\delta} + \kappa(h_2 - r)  )&: h_2 <r<h -h_1,\\
0 &:h -h_1 <r<h,
\end{cases}\label{piecewisetheta}
} 
\eeq
where $\Omega(x)$ is the angular velocity in the core region, and $\kappa(x)$ is the curvature of the velocity profile within the shear layer. From boundary layer theory \cite{schlichting1960boundary}, conservation of radial momentum indicates that radial pressure gradients must balance the centrifugal force from the swirling flow
\beq
\frac{\partial p}{\partial r}=\rho\frac{u_\theta^2}{r},\label{dpdr}
\eeq
where $\rho$ is the density, which is assumed to be constant. Therefore, from (\ref{piecewisetheta}) and (\ref{dpdr}), the approximate pressure profile is
\beq
{\small
p(x,r)=\begin{cases}
p_2+P &: 0<r< h_2,\\
p_1+P &: h_2 < r<h -h_1 ,\\
P &: h -h_1 < r < h ,
\end{cases}\label{piecewisepress}
} 
\eeq
where $p_1(x,r)$ and $p_2(x,r)$ are lengthy functions (which are given in the Supplemental Materials \cite{supp}) found by integrating (\ref{dpdr}) using the approximation (\ref{piecewisetheta}), and $P(x)$ is the pressure in the non-swirling region of flow. Note that $p(x,r)$ is both continuous and once differentiable with respect to $r$.

To model the growth of the shear layer, we modify the model of \citet{benham2018turbulent} to account for the additional effect of azimuthal shear. Thus, we assume that the shear rate $\varepsilon_r$ decays according to
\beq
\frac{U_1+U_2}{2}\frac{d\varepsilon_r}{dx}=-S_c\varepsilon_r^2 \frac{|\mathbf{u}_1-\mathbf{u}_2|}{U_1-U_2}  ,\label{shear2}
\eeq
where $\mathbf{u}_1=(U_1,\,0)$ and $\mathbf{u}_2=(U_2,\Omega h_2)$ are the velocities at either side of the layer, and $S_c$ is an empirical spreading parameter which has been recorded to take values $S_c=0.11-0.18$ \cite{benham2018optimal}. Equation (\ref{shear2}) can be derived from an entrainment argument using a quasi-two-dimensional analogy, which is detailed in the Supplemental Materials \cite{supp}. 


By invoking conservation of mass, axial momentum and angular momentum, we can derive equations describing how $U_1,\,U_2,\,h_1,\,h_2,\,\kappa$ and $P$ evolve. Integrated across the pipe radius, conservation of mass and momentum equations are
\begin{align}
\frac{d}{dx}\int_0^h r u_x \, dr&=0,\label{wallgov1}\\
\rho \frac{d}{dx}\int_0^h r u_x^2 \, dr+\int_0^h r\frac{\partial p}{\partial x}\,dr&=h\tau_{w_r},\label{wallgov2}\\
\rho \frac{d}{dx}\int_0^h r^2 u_\theta u_x\,dr&=0,\label{wallgov3}
\end{align}
where $\tau_{w_r}$ is the shear stress in the axial direction at the wall. 
We assume that the effect of shear stress in the azimuthal direction at the wall is small, so we do not include any stress terms on the right hand side of (\ref{wallgov3}) (though we revise this assumption in our extended model). 
Here, we parameterise the effect of the wall stress in the axial direction using an empirical friction factor $f$, such that
\beq
\tau_{w_r}=-\frac{1}{8} f\rho U_1^2.
\eeq

By choosing a parabolic profile for the azimuthal velocity (\ref{piecewisetheta}) (instead of a linear profile), there is an additional degree of freedom $\Omega(x)$ which is not determined by (\ref{wallgov3}). This is accounted for by imposing conservation of angular momentum within the core region
\beq
\rho\frac{d}{dx}\int_0^{h_2} r^2 u_\theta u_x\,dr=\rho\Omega h_2^2 \frac{d}{dx}\int_0^{h_2} r u_x\,dr 
.\label{innerthetamom}
\eeq
We provide a derivation of (\ref{innerthetamom}), as well as (\ref{wallgov1}) - (\ref{wallgov3}), from the turbulent boundary layer equations in the Supplemental Materials \cite{supp}.

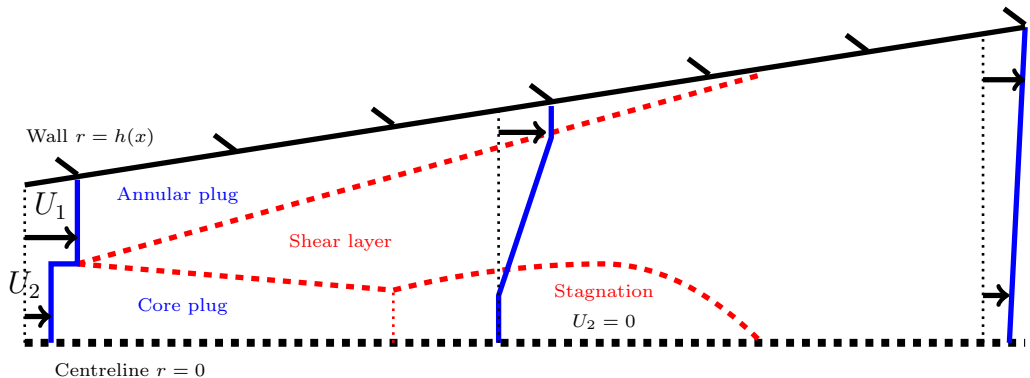
\begin{figure}
\centering
\begin{tikzpicture}[scale=0.7]
\node at (3,-0.5) {\scriptsize Centreline $r=0$};
\node at (2.25,3.9) {\scriptsize Wall $r=h(x)$};
\draw[dashed,line width=3] (1,0) -- (20,0); 
\draw[red,dashed,line width=2] (2,1.5) -- (15,5.1); 
\draw[red,dashed,line width=2] (2,1.5) -- (8,1); 
\draw[red,line width=2,dashed] (8,1) parabola bend (12,1.5) (15,0);
\draw[red,dotted,line width=1] (8,0) -- (8,1); 
\node at (12,0.95) {\scriptsize \color{red} Stagnation};
\node at (7,1.9) {\scriptsize \color{red} Shear layer};
\node at (4,0.7) {\scriptsize \color{blue} Core plug};
\node at (3.9,2.8) {\scriptsize \color{blue} Annular plug};
\node at (12,0.4) {\scriptsize $U_2=0$};
\draw[blue, line width=2] (10, 0) -- (10, 0.9) -- (11,3.9) -- (11,4.5) ; 
\draw[->,line width=2] (10,4) -- (10.9,4); 
\draw[blue, line width=2] (19.7,0) -- (20,6)  ; 
\draw[->,line width=2] (19.2,0.9) -- (19.7,0.9); 
\draw[->,line width=2] (19.2,5) -- (20,5); 
\draw[dotted,line width=1] (19.2,0) -- (19.2,52/9); 
\draw[blue, line width=2] (3/2, 0) -- (3/2,1.5) -- (2,1.5) -- (2,3.1) ; 
\draw[->,line width=2] (1,2) -- (2,2); 
\draw[->,line width=2] (1,0.5) -- (3/2,0.5); 
\draw[dotted,line width=1] (1,0) -- (1,3); 
\draw[dotted,line width=1] (10,0) -- (10,4.5); 
\node at (1.5,2.6) {\large $U_1$};
\node at (1,1.1) {\large $U_2$};
\draw[black,line width=2,-] (1,3) -- (20,6);
\draw[line width=2] (2,3.2) -- (1.6,3.5); 
\draw[line width=2] (5,3.65) -- (4.6,3.95); 
\draw[line width=2] (8,4.15) -- (7.6,4.45); 
\draw[line width=2] (11,4.6) -- (10.6,4.9); 
\draw[line width=2] (14,5.1) -- (13.6,5.4); 
\draw[line width=2] (17,5.55) -- (16.6,5.85); 
\draw[line width=2] (20,6.05) -- (19.6,6.35); 
\node at (3,-1) {};
\end{tikzpicture}
\caption{Schematic diagram of a region of stagnation within a pipe. If the swirl in the core region is sufficiently strong, the flow is sufficiently non-uniform, or if the pipe expansion angle is sufficiently large, the slower core flow can reach zero axial velocity. The aspect ratio here is exaggerated for illustration purposes.  \label{tikz2}}
\end{figure}

{Finally, we neglect energy dissipation within the plug regions, since it is small compared with that near the walls and in the shear layer. 
Therefore, we impose Bernoulli's equation along streamlines in the annular plug region, and for the core region, since pressure varies radially, we only impose Bernoulli's equation along the central non-swirling streamline, such that
\begin{align}
\frac{d}{dx}\lb P+\frac{1}{2} \rho U_1^2 \rb &= 0,\quad\mathrm{or}\quad h_1=0,\label{bern1}\\
\frac{d}{dx}\lb P + p_2(x,0) +\frac{1}{2} \rho U_2^2 \rb &= 0,\quad\mathrm{or}\quad h_2=0,\quad\mathrm{or}\quad U_2= 0 ,\label{bern2}
\end{align}
where we ignore radial velocity components since they are small. Equations (\ref{bern1}) and (\ref{bern2}) are written in this form so that if either plug region is entrained by the shear layer ($h_1=0$ or $h_2=0$), or if the core region slows to zero velocity ($U_2=0$), then the respective Bernoulli's equation no longer holds downstream of that point.
The latter modelling assumption is based on the observation that if the swirl intensity in the core region is sufficiently large, a slow-moving recirculation region can form along the pipe axis, as shown by \citet{lee2005experimental}. A similar recirculation region has been observed for non-swirling flows if the flow is very non-uniform, or if the pipe expansion angle is too large \cite{benham2018optimal}. In our model, we do not attempt to resolve the recirculation within such a region, but approximate it as a stagnant zone with zero axial velocity $U_2=0$ (see Figure \ref{tikz2}).  Later, in Section \ref{cfdcomp}, we show that this approximation shows good agreement with CFD calculations.}

To summarise, our model describes the development of the axial velocity $u_x(x,r)$, given by (\ref{piecewise}), the azimuthal velocity $u_\theta(x,r)$, given by (\ref{piecewisetheta}), and the pressure $p(x,r)$, given by (\ref{piecewisepress}), in a cylindrical pipe. Equations (\ref{shear2})-(\ref{bern2}) govern the variables $U_1,\, U_2,\, \Omega,\, P,\, \delta,\, h_1,\, h_2$, $\varepsilon_r$ and $\kappa$, which are all functions of $x$. We can solve these equations given initial data at $x=0$, numerical values for $f$ and $S_c$, and a pipe shape $h(x)$. For all of the examples we consider here, the shear layer forms at $x=0$, such that initial conditions for $\delta$ and $\varepsilon_r$ are taken as $\delta(0)=0$ and $\varepsilon_r(0)=\infty$ (in practice, we solve for the reciprocal of the shear rate which satisfies $1/\varepsilon_r(0)=0$). Furthermore, pressure is measured with respect to the reference pressure in the annular region at the origin $P(0)=0$, without loss of generality. All other initial conditions form a set of parameters which are displayed in Table \ref{table}, where we write all dimensional parameters in terms of the velocity and length scalings $U_0=U_1(0)$ and $h_0=h(0)$. There is an additional parameter not displayed in Table \ref{table}, which is the non-dimensional swirl number \cite{naughton1997experimental} of the core region, $S_w$. This is defined as the ratio of angular momentum to axial momentum (multiplied by the radius), measured at the inlet, and is given by
\beq
S_w= \left.\frac{\rho \int_0^{h_2} r^2 u_\theta  u_x \, dr }{h_2 \int_0^{h_2} r\lb \rho u_x^2  +(p-p|_{r=h_2}) \rb\, dr }\right|_{x=0}=\frac{2h_2(0)\Omega(0)U_2(0)}{4U_2(0)^2-h_2(0)^2\Omega(0)^2}.\label{tau}
\eeq
Our primary interest is understanding the effect of this swirl number on the flow development, which is discussed in Section \ref{predictions}.

We note that (\ref{tau}) is only well-defined for relatively small amounts of swirl $\Omega(0)<2U_2(0)/h_2(0)$. We also note that (\ref{tau}) depends on $\Omega(0)$ differently, depending on the choice of reference pressure. Here, we have taken the reference pressure as $p|_{r=h_2}=P(0)=0$, which is consistent with \citet{gilchrist2005experimental}.

{
As described earlier, we have also developed an extended version of this model, which accounts for the development of a turbulent boundary layer near the pipe wall and a turbulent symmetry layer near the pipe axis (see Figure \ref{schemext}). Here we briefly summarise the extended model, whilst the complete details are given in the Supplemental Materials \cite{supp}. For the turbulent boundary layer, following \citet{schlichting1960boundary}, we introduce a $1/7$ power law for the axial velocity profile within a region of width $h_b$ near the pipe wall. This velocity profile satisfies the no-slip boundary condition at the wall and matches with the annular plug velocity, $U_1$, at the edge of the boundary layer. The width of the layer, $h_b(x)$, grows according to an equation that derives from integrating conservation of axial momentum over the boundary layer.
The azimuthal velocity profile is given by a similar power law, satisfying the no-slip condition at the wall and matching with the azimuthal velocity at the edge of the boundary layer, which we denote $U_\theta(x)$. The velocity $U_\theta$ is assumed to be $0$ until the shear layer reaches the edge of the boundary layer ($h_1=0$), at which point the swirling flow comes into contact with the boundary layer. In such situations, $U_\theta$ is determined by an equation which derives from conservation of angular momentum within the boundary layer.

\begin{figure}
\centering
\begin{tikzpicture}[scale=0.7]
\node at (3,2.5) {\scriptsize \color{blue} Plug $1$};
\node at (3,0.5) {\scriptsize \color{blue} Plug $2$};
\draw[purple,line width=2,dotted] (1,3) parabola bend (3,3) (20,5); 
\node at (13.5,4.3) {\scriptsize \color{purple} Boundary layer};
\node at (3,-0.5) {\scriptsize Centreline $r=0$};
\node at (2.25,3.8) {\scriptsize Wall $r=h(x)$};
\node at (13,2.2) {\scriptsize \color{red} Shear layer};
\draw[dashed,line width=2] (1,0) -- (20,0); 
\draw[red,dashed,line width=2] (2,1.5) -- (12,3.5); 
\draw[red,dashed,line width=2] (2,1.5) -- (8,0); 
\draw[green,line width=2,dash dot] (8,0) parabola bend (12,0.8) (20,1.);
\node at (14,0.5) {\scriptsize \color{green} Symmetry layer};
\draw[blue, line width=2] (10.4, 0.7) -- (11,3.5) ; 
\draw[line width=2,blue]  (11,3.5) .. controls (11,4.2) .. (10,4.4); 
\draw[line width=2,blue]  (10.35,0) .. controls (10.35,0.5) .. (10.4,0.7); 
\draw[->,line width=2] (10,1.2/2) -- (10.3,1.2/2); 
\draw[->,line width=2] (10,3.6) -- (10.9,3.6); 
\node at (10.4,4.0) {\large $U_1$};
\node at (10.1,1.3) {\large $U_2$};
\draw[blue, line width=2] (19.7,0) -- (19.7,1) -- (20,5.1)  ; 
\draw[line width=2,blue]  (20,5) .. controls (20,5.5) .. (19.2,5.75); 
\draw[->,line width=2] (19.2,0.9) -- (19.7,0.9); 
\draw[->,line width=2] (19.2,5) -- (20,5); 
\node at (19.6,4.3) {\large $U_1$};
\node at (19.3,1.5) {\large $U_2$};
\draw[dotted,line width=1] (19.2,0) -- (19.2,52/9); 
\draw[blue, line width=2] (3/2, 0) -- (3/2,1.5) -- (2,1.5) -- (2,3.1) ; 
\draw[->,line width=2] (1,2) -- (2,2); 
\draw[->,line width=2] (1,0.5) -- (3/2,0.5); 
\draw[dotted,line width=1] (1,0) -- (1,3); 
\draw[dotted,line width=1] (10,0) -- (10,4.5); 
\node at (1.5,2.6) {\large $U_1$};
\node at (1,1.1) {\large $U_2$};
\draw[black,line width=2,-] (1,3) -- (20,6);
\draw[line width=2] (2,3.2) -- (1.6,3.5); 
\draw[line width=2] (5,3.65) -- (4.6,3.95); 
\draw[line width=2] (8,4.15) -- (7.6,4.45); 
\draw[line width=2] (11,4.6) -- (10.6,4.9); 
\draw[line width=2] (14,5.1) -- (13.6,5.4); 
\draw[line width=2] (17,5.55) -- (16.6,5.85); 
\draw[line width=2] (20,6.05) -- (19.6,6.35); 
\draw[<->,line width=1] (6,0.6) -- (6,2.2); 
\draw[<->,line width=1] (6,0.1) -- (6,0.5); 
\draw[<->,line width=1] (6,2.4) -- (6,3); 
\draw[<->,line width=1] (6,3.1) -- (6,3.7); 
\node at (5.5,1.5) {\large $\delta$};
\node at (5.5,0.35) {\large $h_2$};
\node at (5.5,2.6) {\large $h_1$};
\node at (6.5,3.5) {\large $h_b$};
\draw[<->,line width=1] (16,1) -- (16,4); 
\draw[<->,line width=1] (16,0.1) -- (16,0.8); 
\draw[<->,line width=1] (16,4.2) -- (16,5.3); 
\node at (16.5,2.6) {\large $\delta$};
\node at (16.5,0.5) {\large $h_0$};
\node at (16.5,4.8) {\large $h_b$};
\node at (3,-1) {};
\end{tikzpicture}
\caption{Schematic diagram illustrating the boundary layer structure of the extended model, which includes a turbulent boundary layer near the pipe wall and a turbulent symmetry layer near the pipe axis. Axial velocity profiles $u_x$ are illustrated at three locations. \label{schemext}}
\end{figure}
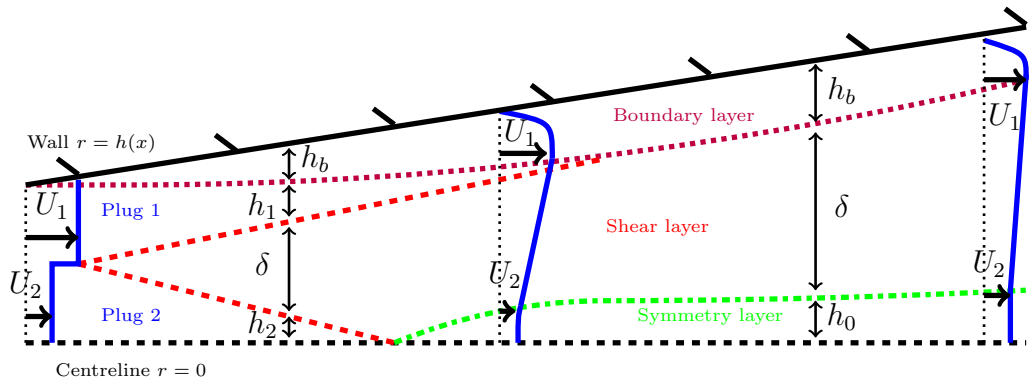

In addition to the boundary layer at the wall, we also introduce a symmetry layer near the pipe axis. The symmetry layer arises in situations where  the shear layer entrains the slower plug region, such that $h_2=0$, and the axial velocity profile (\ref{piecewise}) ceases to satisfy the symmetry condition $\partial u_x /\partial r=0$ at $r=0$. In this case, we introduce a symmetry layer of width $h_0$ near the pipe axis, which has a quadratic axial velocity profile that satisfies the symmetry condition at $r=0$, whilst matching with the velocity at the edge of the shear layer $U_2$. The width of this layer, $h_0(x)$, grows according to an equation that derives from conservation of axial momentum. In summary, the extended model introduces two boundary layers, three extra variables, $h_b$, $U_\theta$ and $h_0$, and three equations which govern these new variables. In the next section, we compare results from both the simple model and the extended model to results from several RANS turbulence models.}

\begin{table}
\centering
\begin{tabular}{|c|c|}
\hline
$U_2(0)/U_0$ & Velocity ratio\\
$h_2(0)/h_0$ & Plug width ratio\\
$h(L)/h_0$ & Expansion ratio\\
$L/h_0$ & Length ratio\\
$S_w$ & Swirl number\\
$S_c$ & Spreading parameter\\
$f$ &  Friction factor\\
\hline
\end{tabular}
\caption{List of the non-dimensional parameters of the problem. We make use of the shorthand $h_0=h(0)$ and $U_0=U_1(0)$. The swirl number $S_w$ is given by Equation (\ref{tau}). \label{table}}
\end{table}


\section{Comparison with RANS turbulence models \label{cfdcomp}}

{In this section we compare results from our simple model and our extended model with those from several steady RANS turbulence models, using the open-source software package \textit{OpenFoam} \cite{weller1998tensorial}.
We use a $k$-$\epsilon$ model \cite{launder1974numerical}, which is probably the most widely used turbulence model, as well as two higher order Reynolds stress turbulence models, which are the $LRR$ model \cite{launder1975progress} and the $SSG$ model \cite{speziale1991modelling}. 
It is reported that these higher order models, which model all components of the Reynolds stresses, can accurately describe swirling flows, whereas two-equation models, such as the $k$-$\epsilon$ model, often fail to do so \cite{jakirlic2002modeling}.}
As examples for our comparison, we first present comparisons of velocity and pressure profiles for co-axial flow in a straight pipe at swirl number $S_w=0.67$. Then we compare predictions of the recirculation region that forms along the axis of an expanding pipe at the same swirl number. Finally, we compare pressure recovery $C_p$ and the outlet kinetic energy flux profile factor $K(L)$ over a range of swirl numbers.

As an example geometry for our first comparison, we choose a straight axisymmetric pipe with non-dimensional length $L/h_0=20$.  The computations are performed using an axisymmetric geometry formed of $30000$ cells, with $300$ elements in the $x$ direction and $100$ elements in the $r$ direction. A convergence check was performed and it was found that this mesh resolution was sufficient for capturing the details of the flow. The inflow velocity ratio is $U_2(0)/U_0=0.5$, the inflow core region radius is given by $h_2(0)=h_1(0)=h_0/2$, and the swirl number is $S_w=0.67$. We use the free-stream boundary conditions for the turbulence variables at the inlet, which are $k=I^2\times 3/2 |\mathbf{u}|^2$ and $\epsilon=0.09 k^{3/2} \ell$, with turbulence intensity $I=5\%$ and mixing length $\ell=0.1 h_0 $. {All of the turbulence parameters in the RANS models are taken as their standard values, which are given by \citet{launder1974numerical, launder1975progress, speziale1991modelling}. }

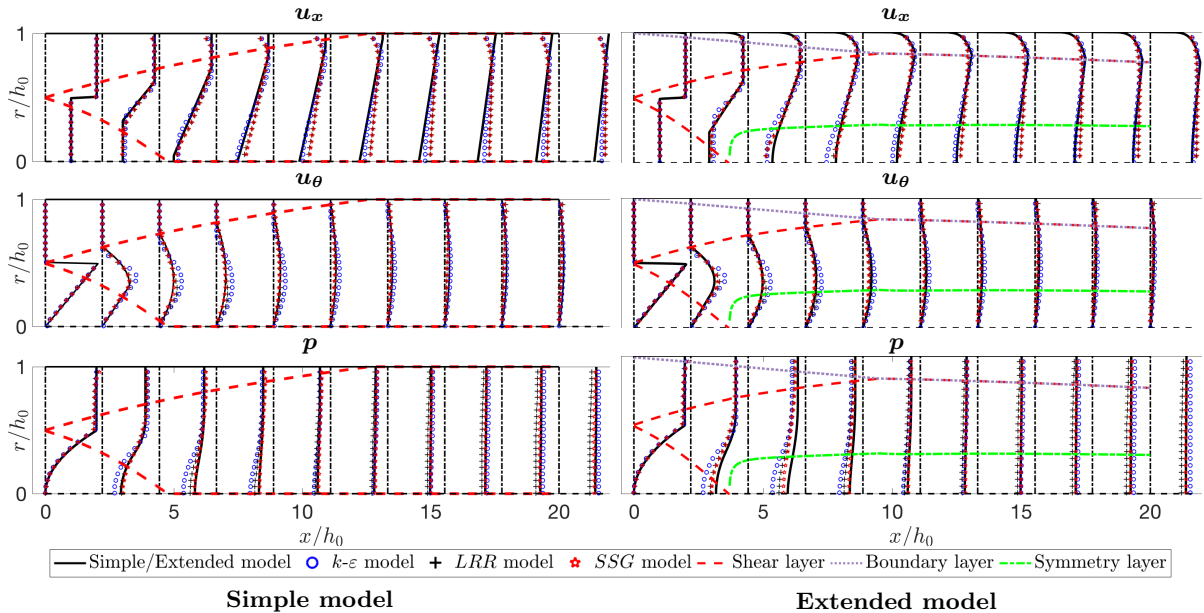
\begin{figure}
\centering
\begin{tikzpicture}
\node at (-1,-2) {\textbf{Simple model}};
\node at (6.8,-2) {\textbf{Extended model}};
\node at (3,-1.5) {\includegraphics[scale=0.4]{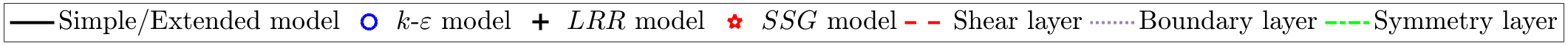}};
\node at (-1.1,4.7) {\includegraphics[scale=0.3]{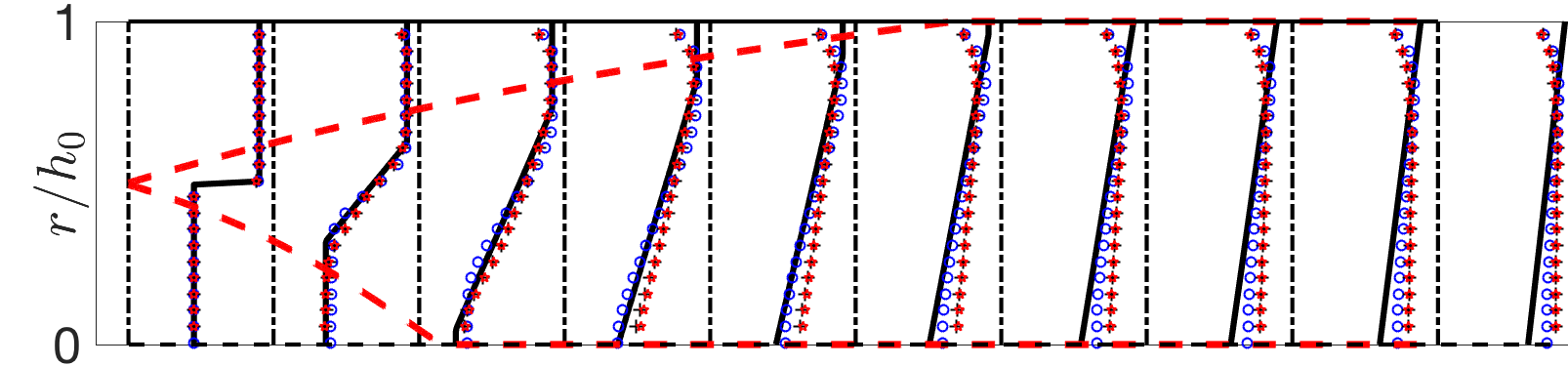}};
\node at (7.0,4.7) {\includegraphics[scale=0.3]{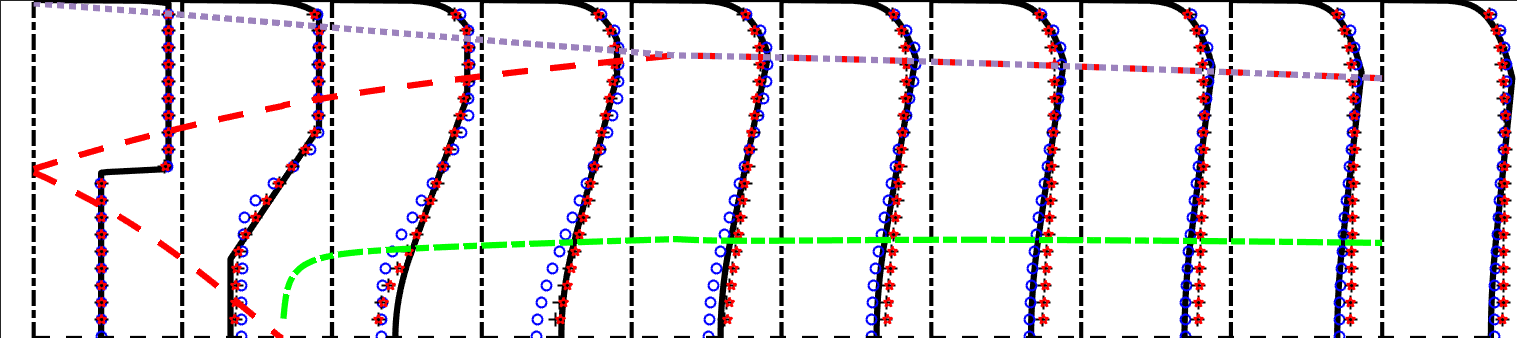}};
\node at (-1.1,2.5) {\includegraphics[scale=0.3]{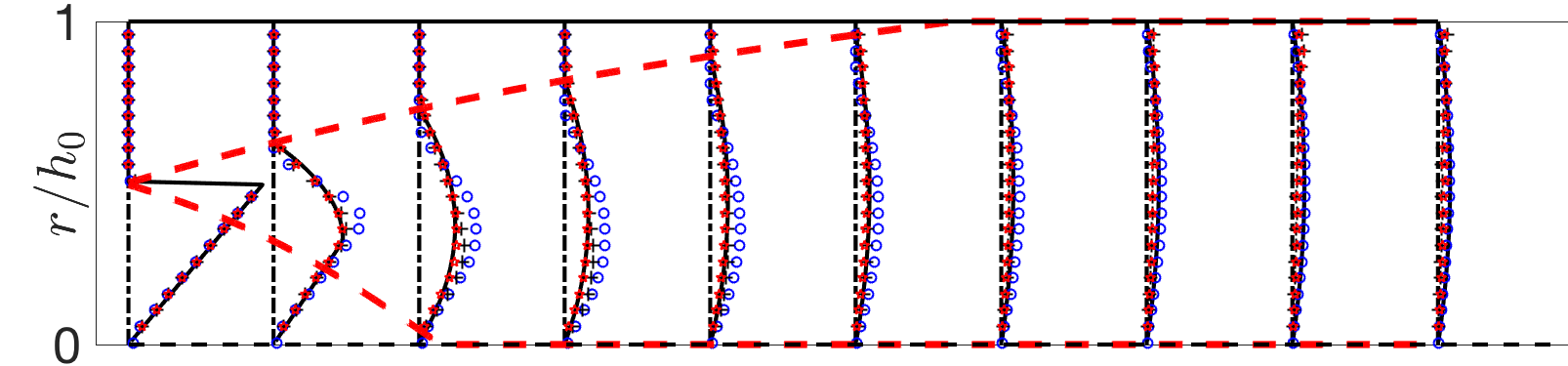}};
\node at (7.0,2.5) {\includegraphics[scale=0.3]{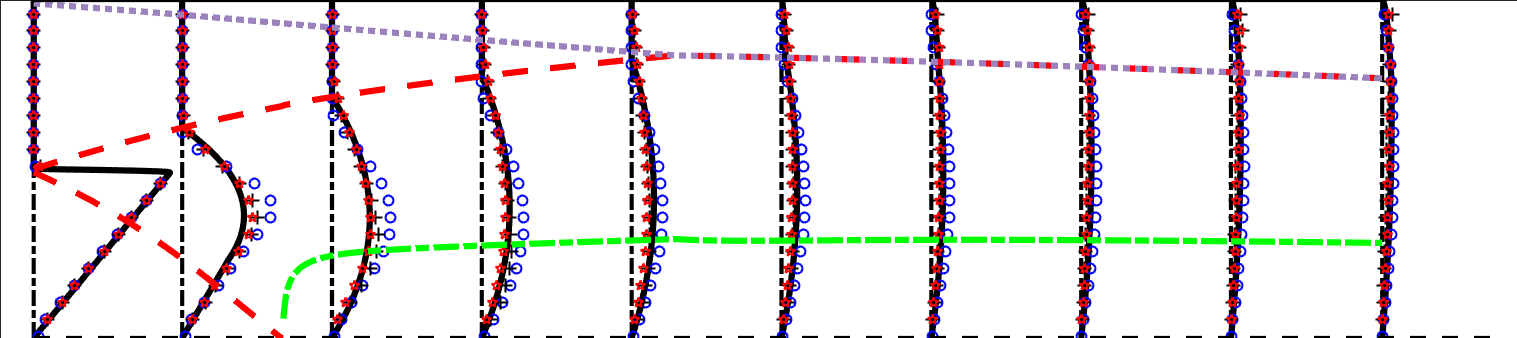}};
\node at (-1.1,0) {\includegraphics[scale=0.3]{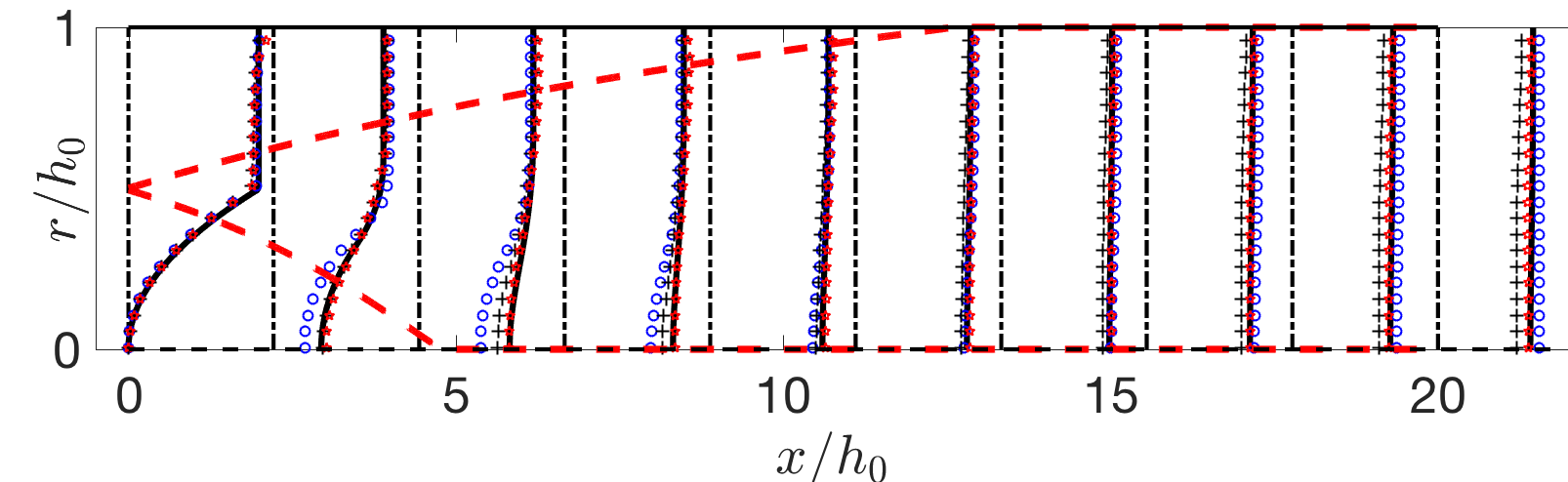}};
\node at (7.0,0) {\includegraphics[scale=0.3]{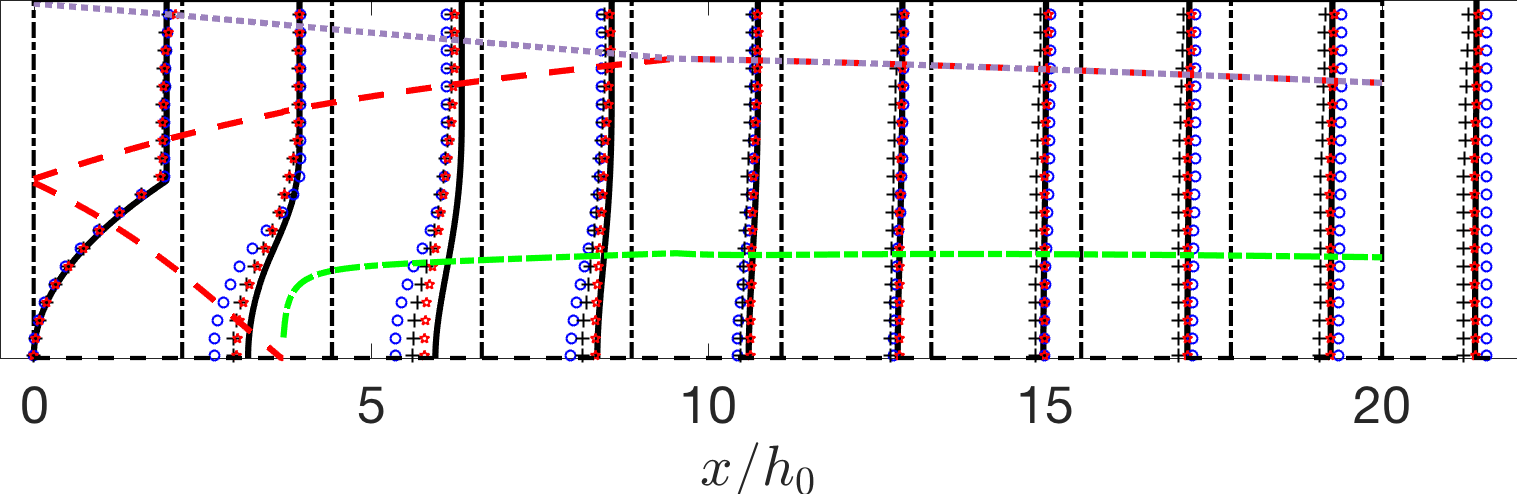}};
\node at (-1,1.4) {$\boldsymbol{p}$};
\node at (6.8,1.4) {$\boldsymbol{p}$};
\node at (-1,3.6) {$\boldsymbol{u_\theta}$};
\node at (6.8,3.6) {$\boldsymbol{u_\theta}$};
\node at (-1,5.8) {$\boldsymbol{u_x}$};
\node at (6.8,5.8) {$\boldsymbol{u_x}$};
\end{tikzpicture}
\caption{Comparison of our simple swirl model and our extended model with {several RANS models} for a swirling slower central flow ($S_w=0.67,\,U_2(0)/U_0=0.5$) mixing with a non-swirling faster outer flow in a straight pipe. Axial profiles are shown at evenly spaced locations for the axial velocity $u_x$, azimuthal velocity $u_\theta$ and pressure $p$. 
\label{modelcfd}}
\end{figure}

If we take typical length and velocity scales as $h_0$ and $U_0$, and we consider the viscosity of water $\nu=10^{-6}\un{m^2/s}$, then the Reynolds number for the inlet flow is $Re=10^6$. For comparison with our simple model and our extended model, we estimate the friction factor with the Blasius relationship \cite{blasius1913ahnlichkeitsgesetz, mckeon2005new} for flow in smooth pipes $f=0.316 Re^{-1/4}$, giving a value of $f=0.01$. For the spreading parameter $S_c$, we find that {$S_c=0.13$} gives the best agreement, which falls within the range of reported values \cite{benham2018optimal}. {In the Supplemental Materials \cite{supp} we also compare our model to other geometries and at different swirl numbers, and we find that {$S_c=0.13$} is consistently an appropriate value for a good fit. }

In Figure \ref{modelcfd} we compare profiles of axial velocity $u_x$, azimuthal velocity $u_\theta$ and pressure $p$ at evenly spaced locations along the pipe. 
The comparison shows that there is good agreement between the RANS models and both our simple model and our extended model, though the extended model has better agreement overall. 
{The mean relative errors between our simple model and the $k$-$\epsilon$, $LRR$ and $SSG$ models are $6.0\%$, $7.9\%$ and $5.9\%$, compared to $5.0\%$, $5.2\%$ and $4.0\%$ for the extended model. 
The $k$-$\epsilon$ model appears to slightly under-predict the growth rate of the shear layer and the development of the swirl profile (compared to the higher order models).
For both the simple model and the extended model, the width of the shear layer, the maximum and minimum velocities and the parabolic swirl profile (and consequently the pressure) in the shear layer match closely to the RANS models, especially the $SSG$ model}. Since our simple model does not explicitly resolve the development of boundary layers, the comparison is less close near the pipe walls. 
However, pressure variations in the axial direction, which are controlled by $f$, agree well (see Figure \ref{modelcfd}), indicating that the effect of the boundary layers is captured by the simple model using the friction factor parameterisation. 
Furthermore, we can see that the comparison is also less close downstream, near $r=0$, since the simple model does not satisfy the symmetry condition, $\partial u_x /\partial r=0$ at $r=0$, when the shear layer comes into contact with the pipe axis. 
In the case of the extended model, the development of the turbulent boundary layer, which is shown by a purple dotted line, matches closely with all the RANS models, as does the development of the turbulent symmetry layer, which is shown by a green dash-dotted line. 

The extended model has roughly the same computational cost as the original swirl model, though it contains more variables and equations, and it is perhaps more difficult to implement. Therefore, the original model serves as a convenient tool for benchmark predictions, whilst the extended model is appropriate for more detailed and accurate calculations of flow development, which take longer to implement.

\begin{figure}
\centering
\begin{overpic}[width=0.49\textwidth]{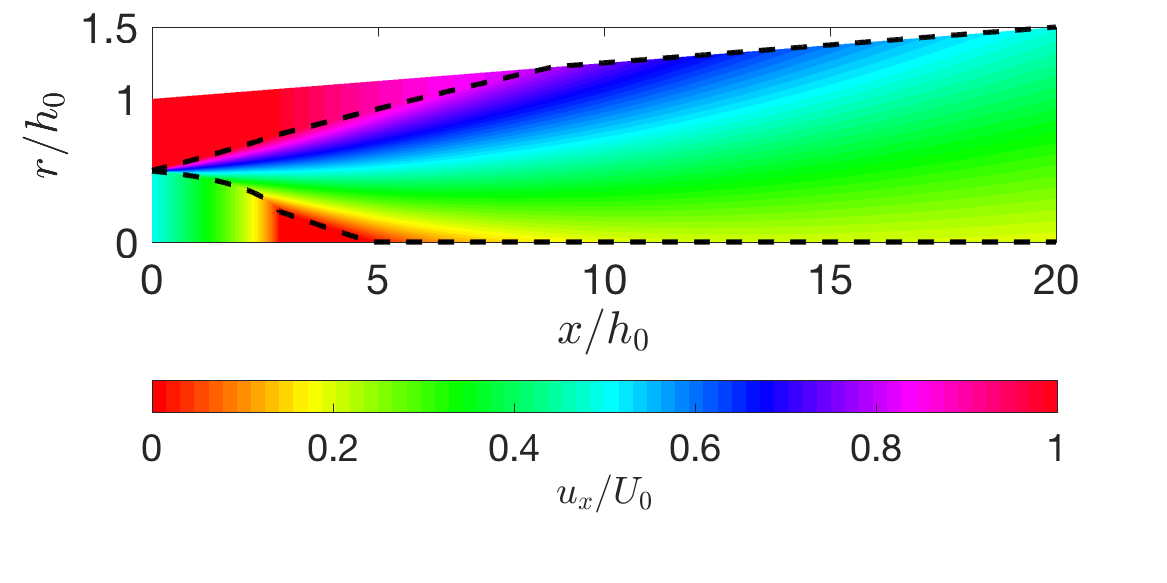}
\put (0,50) {(a)}
\put (35,50) {\textbf{Simple model}}
\end{overpic}
\begin{overpic}[width=0.49\textwidth]{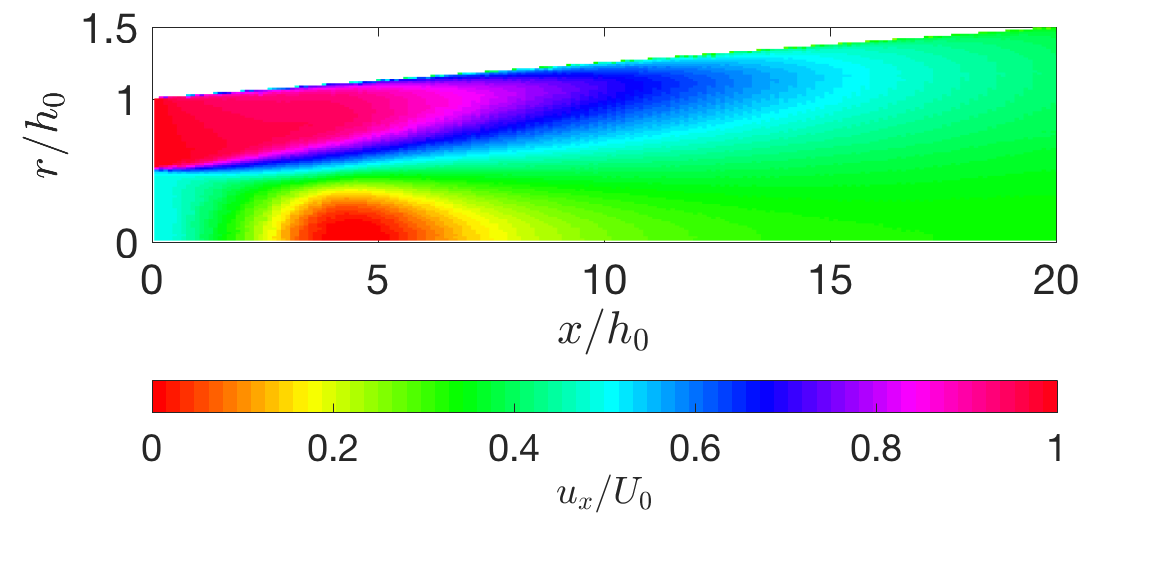}
\put (0,50) {(b)}
\put (40,50) {\textbf{$SSG$ model}}
\end{overpic}\\
\begin{overpic}[width=0.49\textwidth]{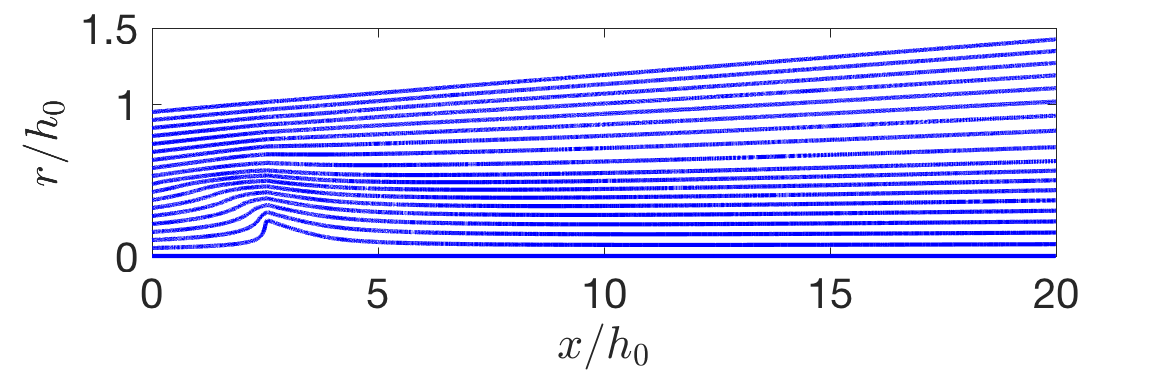}
\put (0,35) {(c)}
\end{overpic}
\begin{overpic}[width=0.49\textwidth]{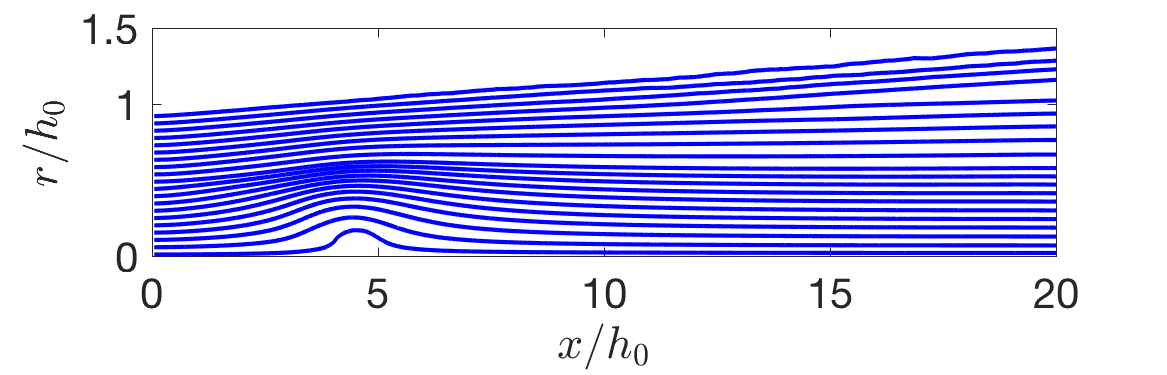}
\put (0,35) {(d)}
\end{overpic}
\caption{Comparison of the stagnation region in an expanding pipe with expansion angle $1.4^\circ$, displaying velocity colour plots and streamlines, calculated using our simple model and an $SSG$ model. The swirl number is $S_w=0.67$ and the inlet profile is given by $U_2(0)/U_0=0.5,\,h_1(0)=h_2(0)=h_0/2$. \label{stagcomp}}
\end{figure}

Next, we compare predictions of the stagnation region which can form along the pipe axis. In Figure \ref{stagcomp} an example of a stagnation region is shown for an expanding pipe with expansion angle $1.4^\circ$ and non-dimensional length $L/h_0=20$. The swirl number is $S_w=0.67$ and the inlet profile is given by $U_2(0)/U_0=0.5,\,h_1(0)=h_2(0)=h_0/2$. {Velocity colour plots and streamlines are displayed for both the simple model and the $SSG$ model, where in the case of the streamlines, the radial velocity of our simple model is calculated from the axial velocity,}
\beq
u_r=-\frac{1}{r}\int_0^r r \frac{\partial u_x}{\partial x}\,dr.
\eeq
The size and position of the stagnation region show close agreement, indicating that the simple model has good predictive capabilities. 
Various studies \cite{chigier1967experimental, escudier1985recirculation, champagne2000experiments, nayeri2000investigation} relate this region to the phenomenon of vortex breakdown. 
In our model, it can be explained by Equations (\ref{dpdr}), (\ref{shear2}) and (\ref{bern2}). 
Due to (\ref{dpdr}), larger amounts of swirl induce a lower pressure along the pipe axis $r=0$. As the shear layer spreads out due to (\ref{shear2}), the azimuthal velocity profile (\ref{piecewisetheta}) flattens, causing a rise in pressure along the pipe axis. Hence, from Bernoulli's equation (\ref{bern2}), this rise in the pressure causes the speed of the core region to drop.
If the effect is sufficiently pronounced then the core region will stagnate completely.

As a final comparison between our simple model and CFD calculations, we make use of two commonly used measures of flow performance: the pressure recovery coefficient $C_p$ and the kinetic energy flux profile factor $K$ \cite{blevins1984applied}. The pressure recovery coefficient $C_p$ is a measure of the amount of kinetic energy in a flow which is converted into static pressure. There are various ways to define $C_p$, but we use the so-called mass-averaged pressure recovery \cite{filipenco1998effects}, which is defined as 
\begin{equation}
C_p=\frac{ \int_0^h r u_x p \,dr |_{x=L} - \int_0^h r u_x p \,dr |_{x=0}}{\int_0^h \frac{1}{2}\rho r u_x |\mathbf{u}|^2  \,dr |_{x=0} },\label{Cpdef}
\end{equation}
and takes values between $-\infty$ and $1$, where $C_p=1$ if all the inlet kinetic energy is converted into static pressure. The kinetic energy flux profile factor $K(x)$ is a measure of how non-uniform the axial velocity profile of a flow is, at a given position $x$,
defined as
\begin{equation}
K(x)=\frac{\frac{2}{h^2}\int_0^h r u_x^3 \,dr}{\lb\frac{2}{h^2}\int_0^h r u_x  \,dr\rb^3},
\end{equation}
and takes values between $1$ and $\infty$, where $K=1$ corresponds to uniform flow. 

\begin{figure}
\centering
\begin{overpic}[width=0.49\textwidth]{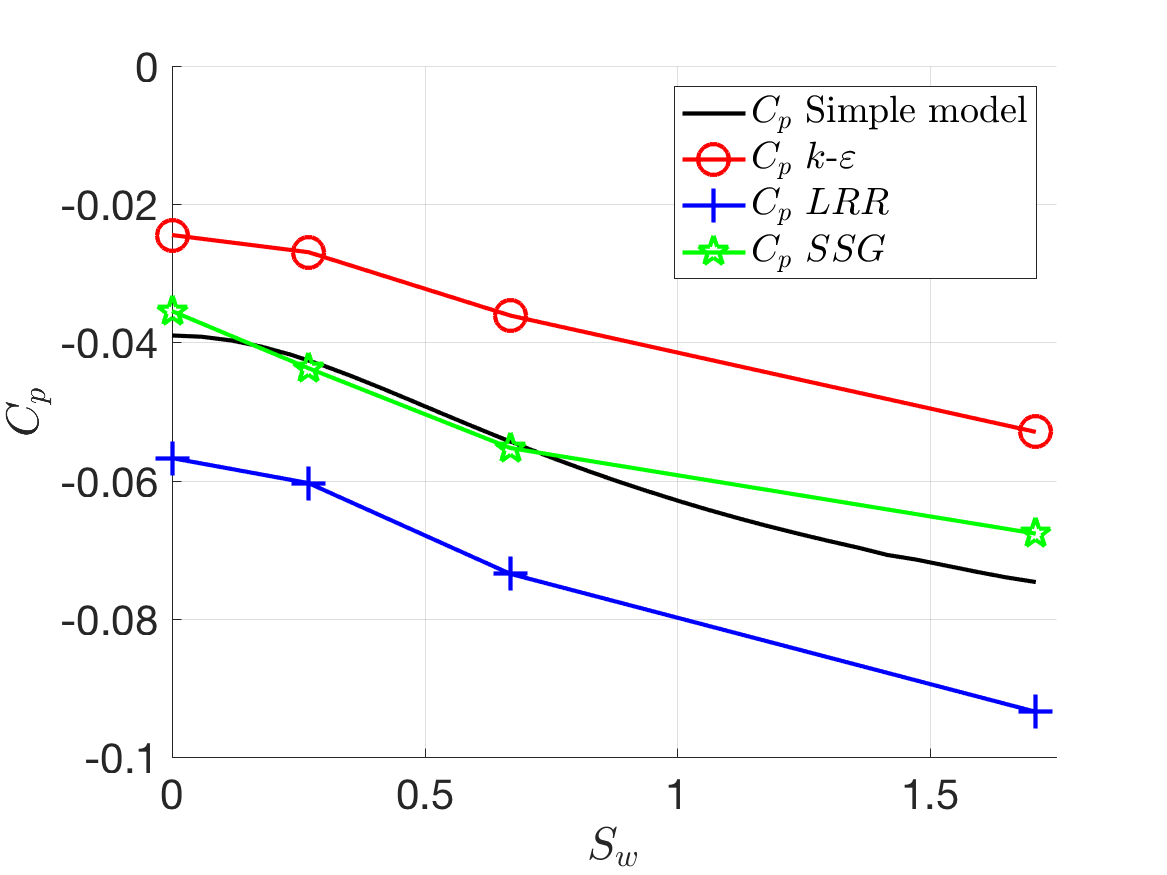}
\put (0,70) {(a)}
\end{overpic}
\begin{overpic}[width=0.49\textwidth]{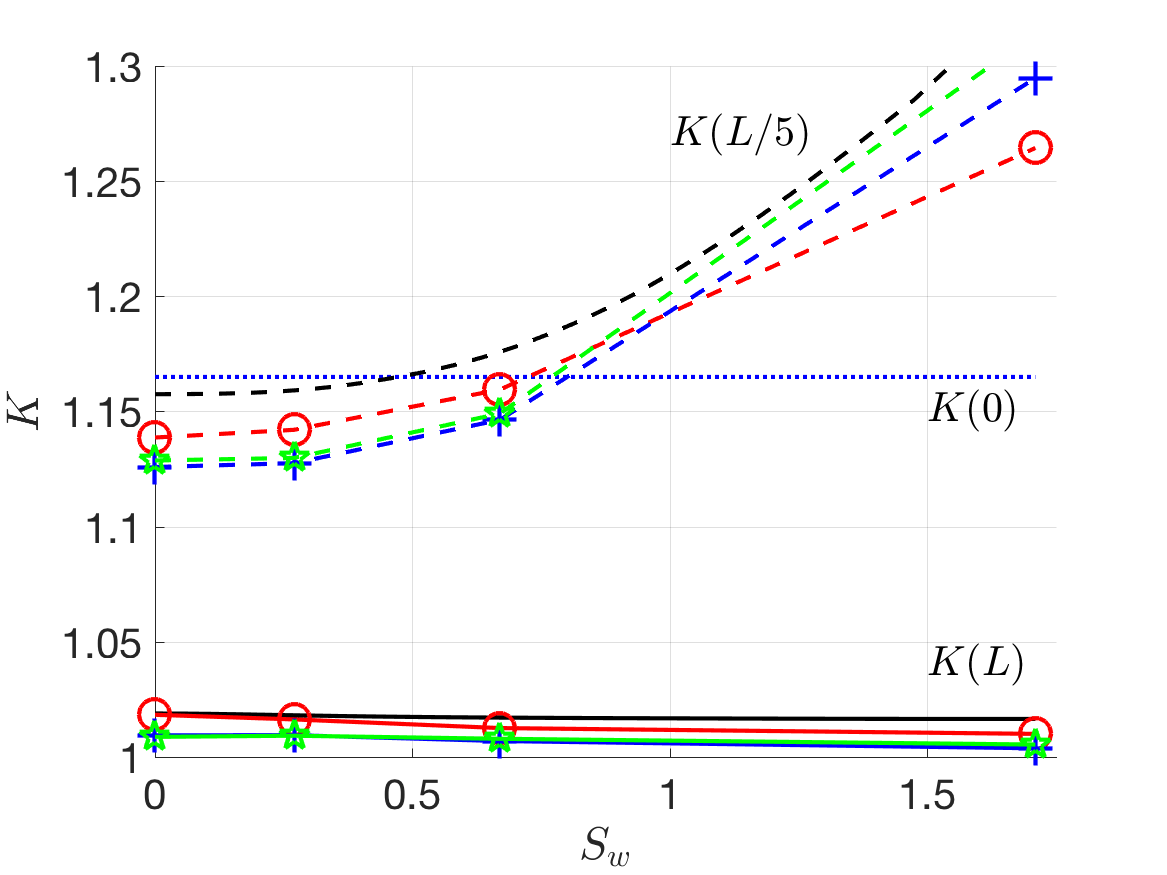}
\put (0,70) {(b)}
\end{overpic}
\caption{Comparison of the pressure recovery coefficient $C_p$ and the kinetic energy flux profile factor $K$ calculated using both our simple model and {several RANS models}, where we vary the swirl number $S_w$. The kinetic energy flux profile factor is measured at $x=0,\,L/5$ and $L$. In these calculations, we consider the same flow geometry as in Figure \ref{modelcfd}. \label{CpK}}
\end{figure}

For the straight pipe in Figure \ref{modelcfd} {we compare calculations of both $C_p$ and $K$ (measured at $x = 0, L/5$ and $L$) using our simple swirl model and the RANS models for values of the swirl number between $S_w = 0$ and $S_w = 1.71$. The results are plotted in Figure \ref{CpK}. Overall, there is good agreement, and both models show that increased swirl has the effect of reducing pressure recovery, whilst making the flow more non-uniform in the near-field (at $x = L/5$), and then more uniform downstream (at $x = L$). Large swirl makes the flow more non-uniform in the near-field due to its distorting effect on the velocity profile \cite{so1967vortex}. However, it makes the flow more uniform downstream due to increased shear layer growth rates \cite{gilchrist2005experimental} and an overall enhanced mixing effect. These conclusions can be interpreted in the following way: Whilst swirl may produce a more uniform outflow, it distorts the flow in the near-field, resulting in greater energy dissipation, which manifests in a loss of pressure.}

In the Supplemental Materials \cite{supp} we also compare our model to some experimental data from other studies of an unconfined swirling jet, finding good agreement. 
This and all of the previous comparisons give us confidence that our simple model has good predictive capabilities, whilst being computationally cheap (e.g. using finite differences with $500$ discretisation points, computation time is of the order of $1\un{s}$ on a laptop computer). Therefore it is useful for examining the effect of swirl in a wide range of flow situations and exploring design spaces. 

\section{Model predictions \label{predictions}}

In this section we use our simple model to analyse the effect of swirl on different flow situations. First, we look at the effect on the behaviour of the shear layer, determining the distance downstream at which the shear layer has reached across the entire pipe (this information is important for problems involving momentum transfer). Secondly, we establish which swirl numbers and pipe geometries result in a stagnation region forming along the pipe axis. Finally, we use our simple model to optimise the shape of a pipe for different swirl numbers, using both $C_p$ and $K(L)$ as design objectives.

\subsection{Shear layer behaviour}

\begin{figure}
\centering
\vspace{0.5cm}
\begin{overpic}[width=0.49\textwidth]{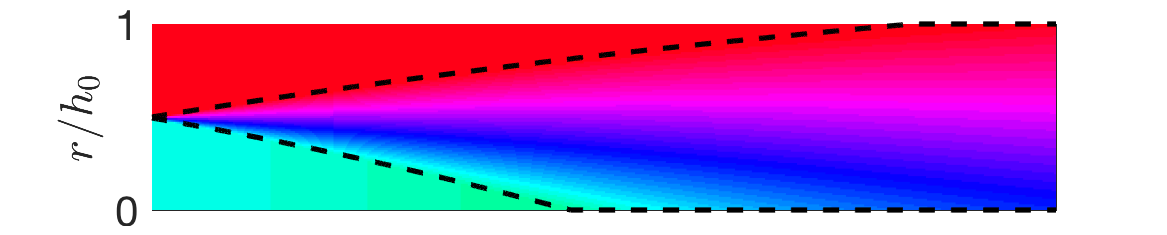}
\put (-5,20) {(a)}
\put(-1,0){\rotatebox{90}{\scriptsize $\boldsymbol{S_w=0.27}$}}
\put(35,0){\color{black}\vector(1,0){15}}
\put(27,0){\color{black}\vector(-1,0){15}}
\put (28,-1.5) {\small$L_2$}
\put(40,20){\color{black}\vector(1,0){20}}
\put(32,20){\color{black}\vector(-1,0){20}}
\put (33,19.5) {\small$L_1$}
\end{overpic}
\begin{overpic}[width=0.49\textwidth]{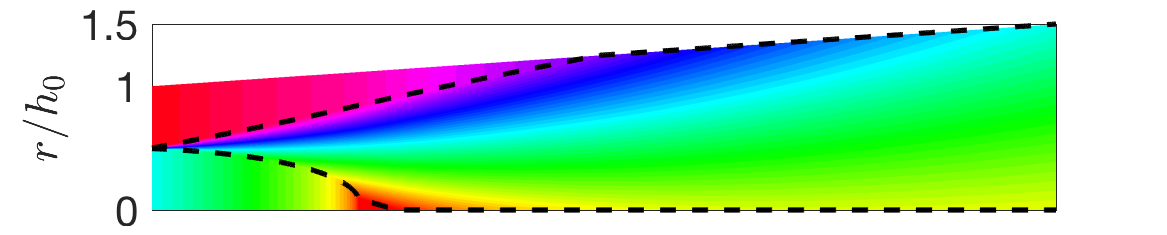}
\put (0,20) {(b)}
\end{overpic}
\vspace{0.2cm}\\
\begin{overpic}[width=0.49\textwidth]{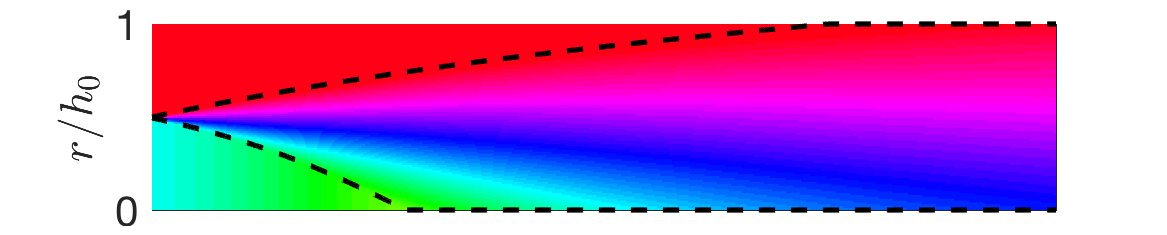}
\put(-1,-2){\rotatebox{90}{\scriptsize  $\boldsymbol{S_w=0.67}$}}
\put (-5,17) {(c)}
\end{overpic}
\begin{overpic}[width=0.49\textwidth]{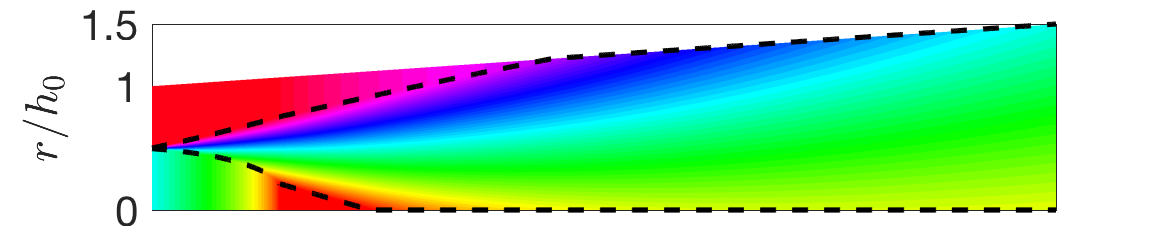}
\put (0,17) {(d)}
\end{overpic}\vspace{0.2cm}\\
\begin{overpic}[width=0.49\textwidth]{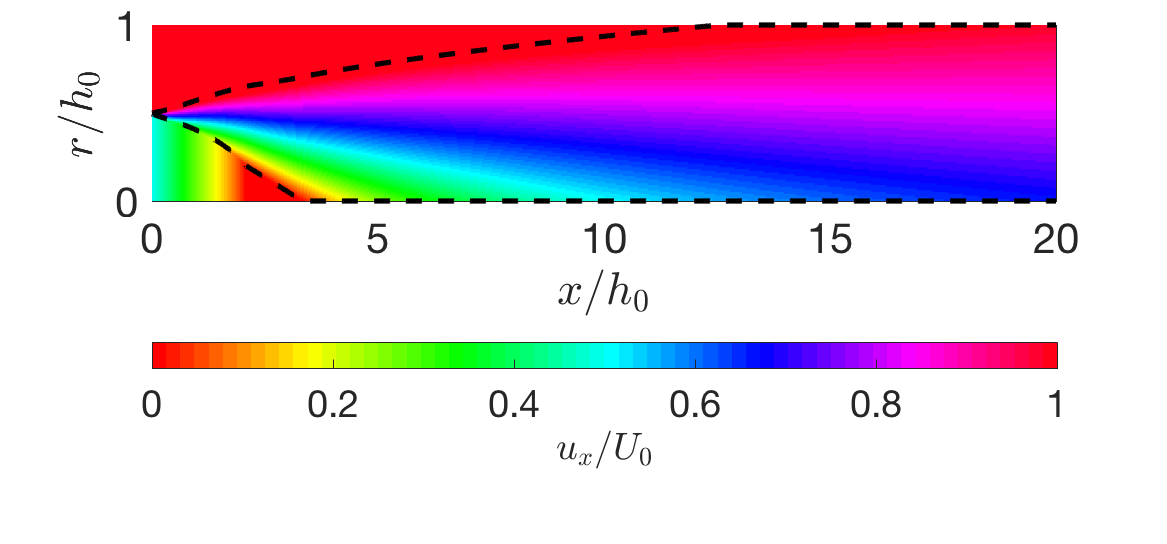}
\put(-1,22){\rotatebox{90}{\scriptsize  $\boldsymbol{S_w=1.71}$}}
\put (-5,43) {(e)}
\end{overpic}
\begin{overpic}[width=0.49\textwidth]{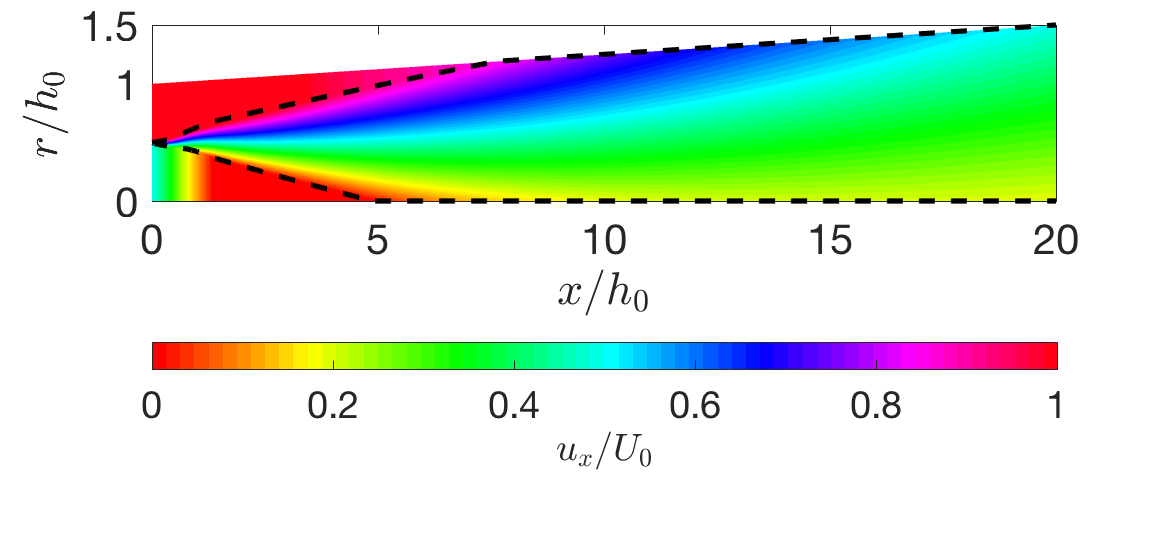}
\put (0,45) {(f)}
\end{overpic}
\begin{overpic}[width=0.49\textwidth]{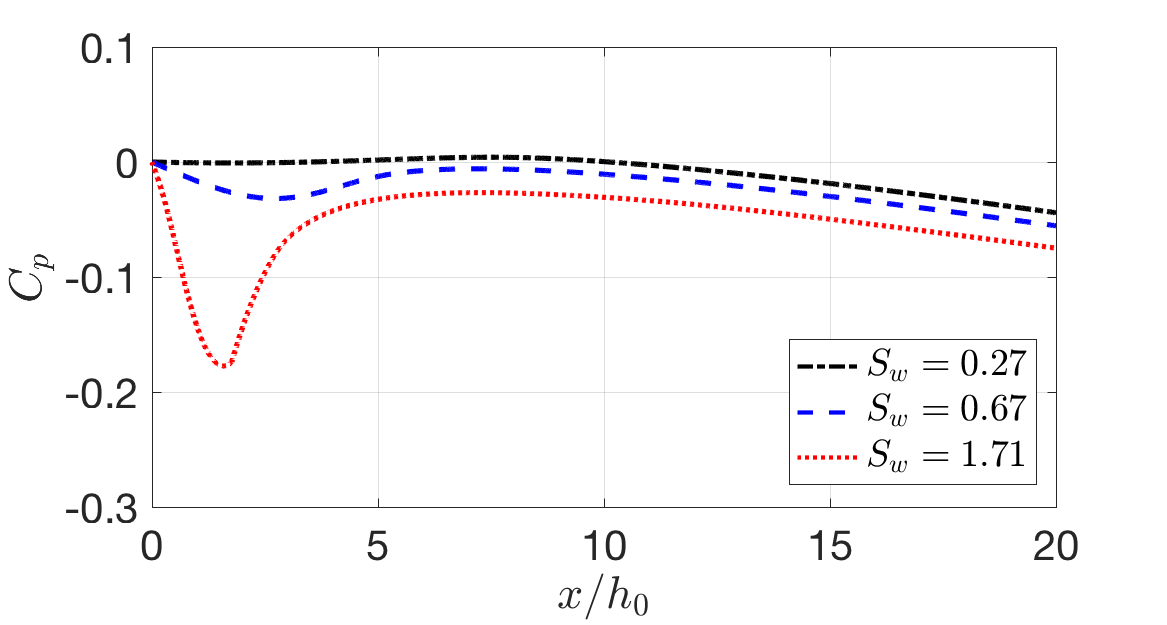}
\put(50,52){\textbf{$C_p$}}
\put (40,-5) {\bf Straight}
\put (0,55) {(g)}
\end{overpic}
\begin{overpic}[width=0.49\textwidth]{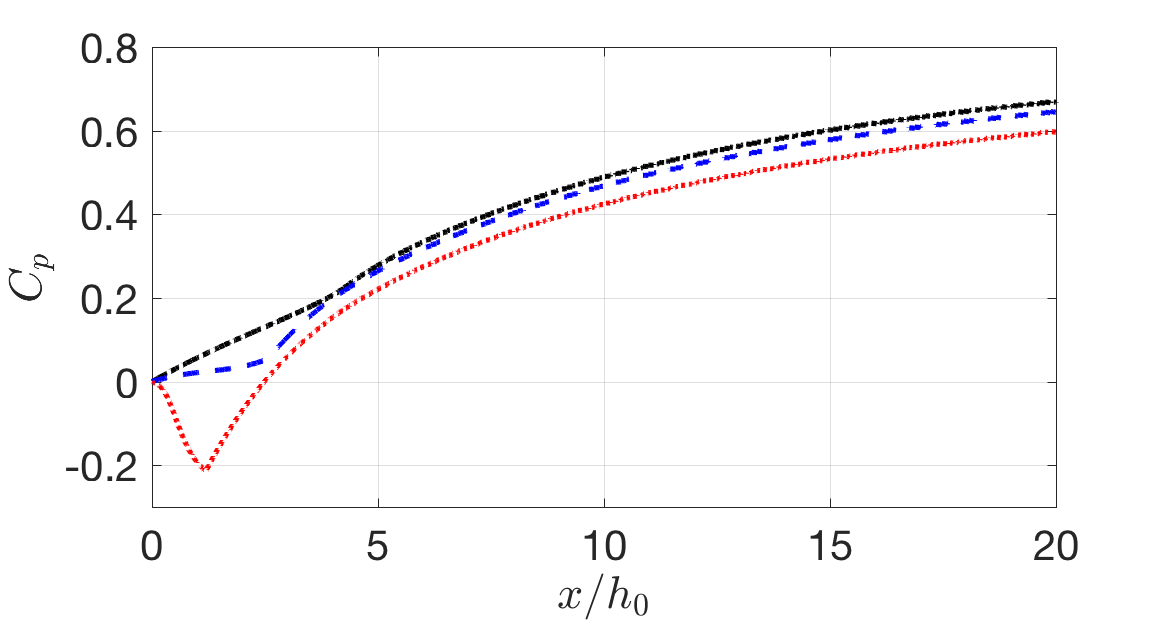}
\put(50,52){\textbf{$C_p$}}
\put (40,-5) {\bf Widening}
\put (0,55) {(h)}
\end{overpic}
\vspace{0.2cm}
\caption{The effect of increasing the swirl number $S_w$ between $0.27$ and $1.71$ on the axial velocity $u_x$ in the case of a straight pipe (a, c, e, g) and a widening pipe (b, d, f, h). The position of the shear layer is indicated in dashed black lines. Pressure recovery $C_p$, given by (\ref{Cpdef}), is plotted as a function of $x/h_0$ for each case. The distances, $L_1$ and $L_2$, at which the annular and core regions are entrained by the shear layer, respectively, are illustrated in (a). The inlet profile is given by $U_2(0)/U_0=0.5$, $h_1(0)=h_2(0)=h_0/2$. \label{strwid}}
\end{figure}
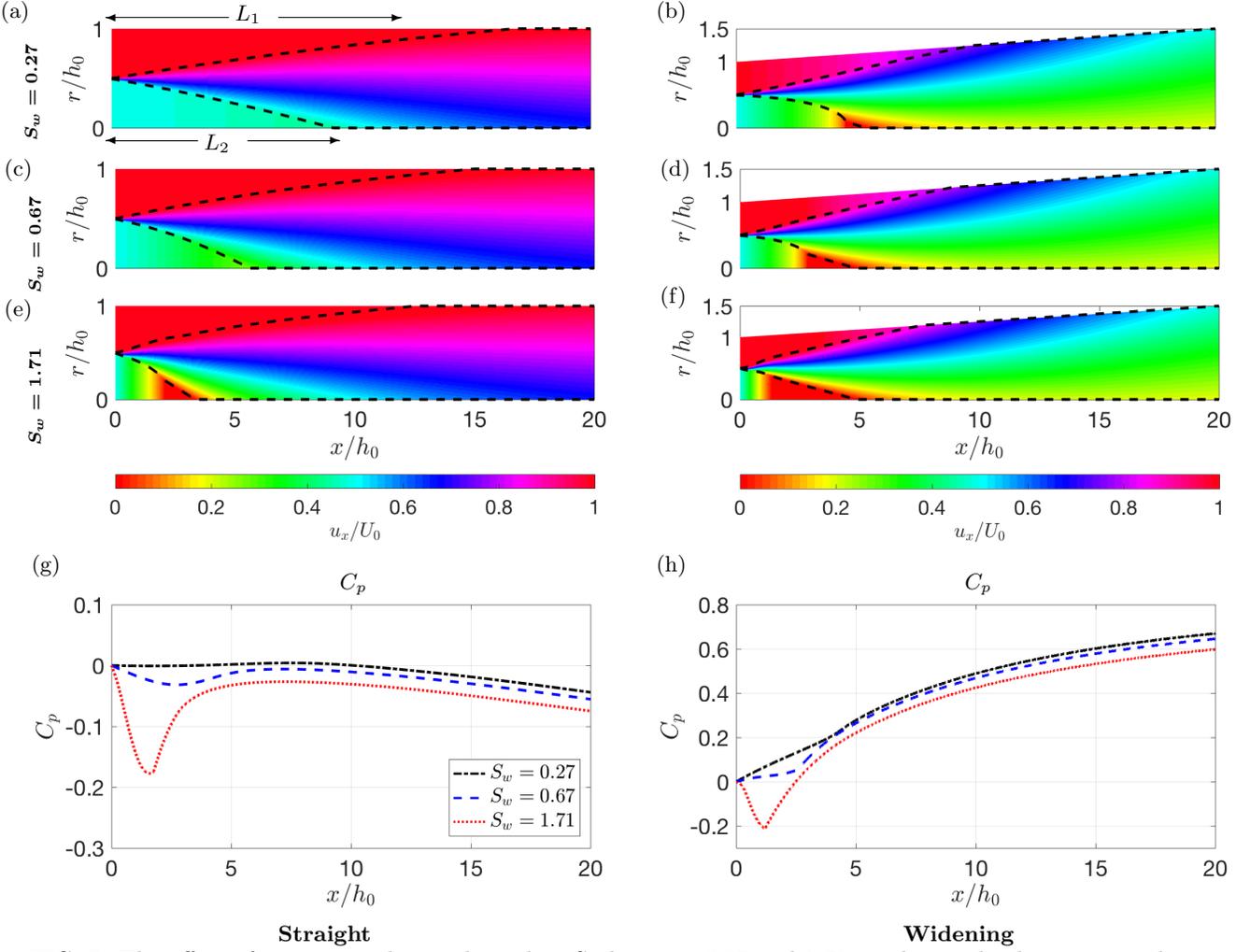

One useful way to characterise the shear layer behaviour is by looking at the growth rate. However, since shear layer growth rates are not constant, this requires choosing a length scale over which one can take an average of the growth rate. The choice of this length scale is slightly arbitrary and will have a significant effect on growth rate measurements \cite{gilchrist2005experimental}. In situations where the flow is confined, there is a more natural way to measure the shear layer growth rate, which is the distance it takes for the shear layer to reach across the pipe. Since the shear layer has an inner and an outer edge, which are not necessarily symmetric, it is useful to consider where each of these edges reaches their respective boundaries. Thus, the outer edge of the shear layer must reach the pipe wall $r=h$ at some distance $x=L_1$, whereas the inner edge of the shear layer reaches the pipe axis $r=0$ at $x=L_2$ (see Figure \ref{strwid} (a)). 

In Figure \ref{strwid} we display colour plots of non-dimensional axial velocity $u_x/U_0$, and plots of the pressure recovery coefficient $C_p$, in both a straight pipe and a widening pipe (with expansion angle $1.4^\circ$), for different values of the swirl number. For this example we choose a pipe with non-dimensional length $L/h_0=20$ and an inflow which is defined by $U_2(0)/U_0=0.5$ and $h_1(0)=h_2(0)=h_0/2$.
By observation, it is apparent that the shear layer growth rate ${d}\delta/{d}x$ varies with $x$, and increases on average with the swirl number. 

The corresponding values of $L_1$ and $L_2$ are plotted in Figure \ref{Lstag} (a). We can clearly see that for both the straight pipe and the widening pipe, $L_1$ and $L_2$ decrease as the swirl number increases, confirming that shear layer growth rates are enhanced by swirl. We can also see that the values of $L_1$ and $L_2$ are smaller for the widening case, indicating that pipe expansion is another mechanism for enhanced shear layer growth rates.

\subsection{Stagnation region}

As is discussed by \citet{lee2005experimental}, flows with large swirl numbers have the tendency to form a region of slowly moving recirculation. Here, we use our simple model to characterise the onset and size of this region by approximating it as a stagnant zone with zero velocity, as described earlier. 

In the examples in Figure \ref{strwid} we can see a stagnation region forming for swirl number $S_w=1.71$ in the straight pipe case, and for all displayed values of $S_w$ in the widening pipe case. In each case where the stagnation region forms, it then closes again some distance downstream. The results from our simple model indicate that the size of the stagnation region increases with $S_w$. Furthermore, it is evident that expanding pipes are more susceptible to stagnation than straight pipes. In the plots of pressure recovery $C_p$, we see a significant dip in pressure recovery where the flow stagnates. This is due to the fact that as the flow stagnates, the axial velocity profile becomes very non-uniform, resulting in an increase in kinetic energy flux. Therefore, $C_p$, which is a measure of how much inflow kinetic energy is converted to static pressure, must decrease \cite{blevins1984applied}.

\begin{figure}
\centering
\begin{overpic}[width=0.48\textwidth]{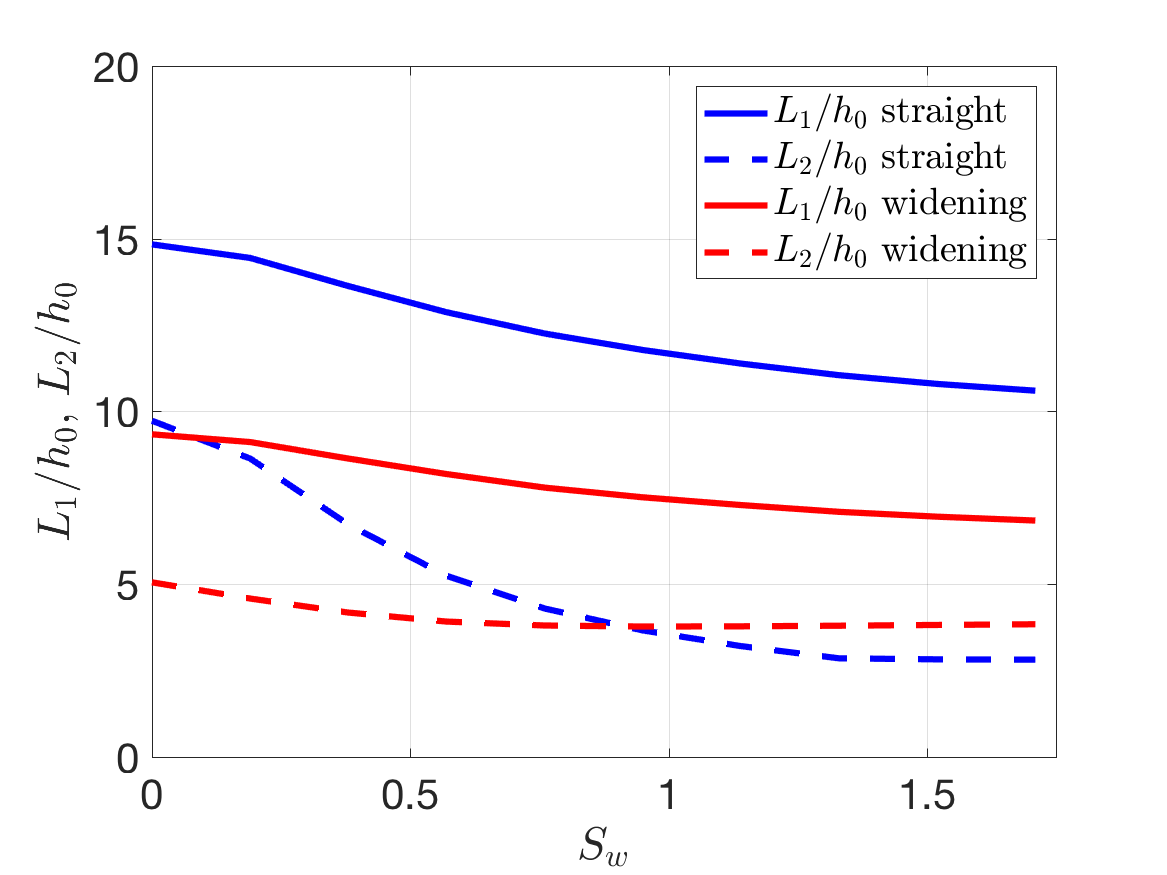}
\put (0,70) {(a)}
\end{overpic}
\begin{tikzpicture}
\node at (1,1) {\includegraphics[width=0.48\textwidth]{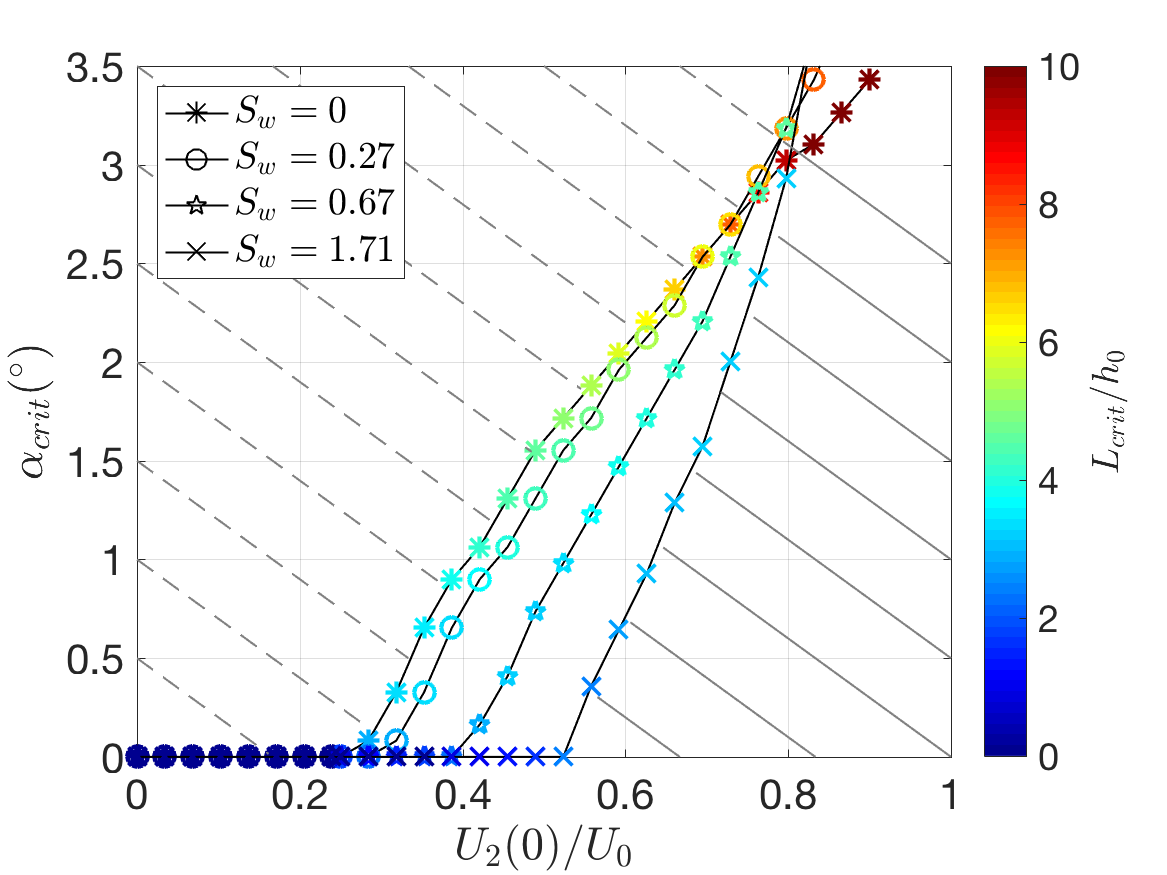}};
\node at (-0.3,1.3) {\scriptsize \textbf{Stagnation}};
\node at (-3.3,3.7) {(b)};
\node at (2.5,0.2) {\scriptsize  \textbf{No}};
\node at (2.5,-0.1) {\scriptsize  \textbf{Stagnation}};
\end{tikzpicture}
\caption{Effect of swirl on flow development properties. (a) $L_1/h_0$ and $L_2/h_0$ are the non-dimensional distances downstream at which the slower core and faster annular regions are completely entrained by the shear layer (see Figure \ref{strwid} (a)). The inflow for this example is defined by $U_2(0)/U_0=0.5$ and $h_1(0)=h_2(0)=h_0/2$. (b) The minimum diffuser angle $\alpha_{crit}$ at which flow in a pipe with linearly expanding walls stagnates, for different values of the inlet velocity ratio $U_2(0)/U_0$ and swirl number $S_w$. The non-dimensional distance downstream at which stagnation first occurs $L_{crit}/h_0$ is indicated by the colour bar (which is saturated above $L_{crit}/h_0>10$).  \label{Lstag}}
\end{figure}
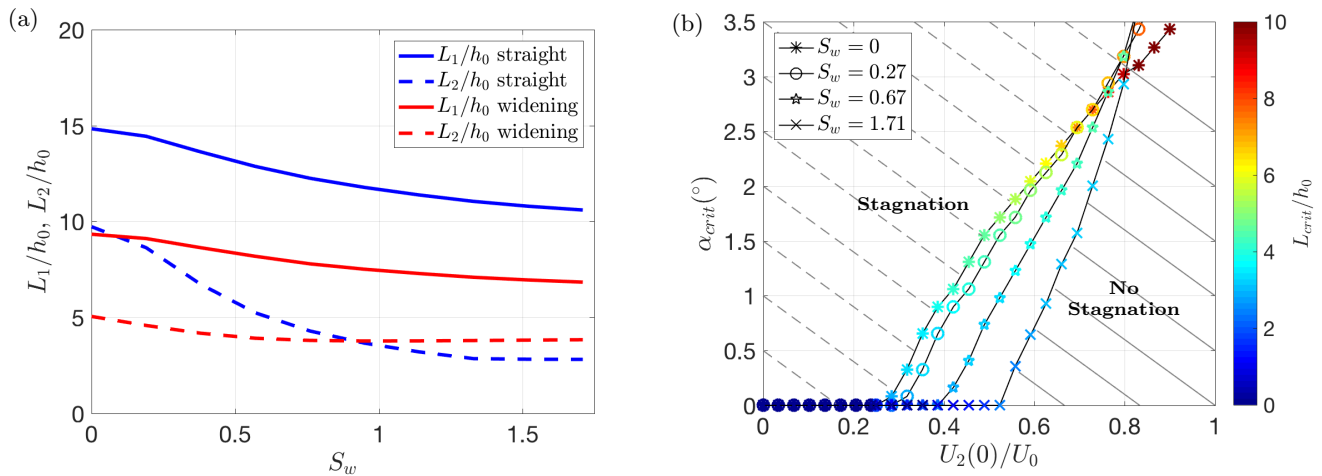

In Figure \ref{Lstag} we display a plot indicating the criteria for flow stagnation in a pipe of constant expansion angle, for different values of the swirl number. On the $x$-axis we vary the inlet velocity ratio $U_2(0)/U_0$, whilst on the $y$-axis we plot the expansion angle $\alpha_{crit}$ above which stagnation occurs. In each case the radius of the core region is such that $h_1(0)=h_2(0)=h_0/2$.
Regions of the parameter space where a stagnation region appears are indicated by dashed lines. 
Furthermore, the non-dimensional distance downstream at which the stagnation region forms, which we denote $L_{crit}/h_0$, is indicated by the colour of the data points that demarcate the boundary between stagnation and no-stagnation. We can see that the flow is less susceptible to stagnation for velocity ratios closer to $1$, since the critical expansion angle $\alpha_{crit}$ for which stagnation occurs is larger, and the critical distance $L_{crit}/h_0$ is also larger. 

The effect of increasing the swirl number is interesting and possibly counter-intuitive. For velocity ratios less than around $U_2(0)/U_0\approx 0.7$, increasing $S_w$ causes a reduction in the critical expansion angle, thereby increasing the range of parameter values for which stagnation occurs. This is expected and consistent with the findings of \citet{lee2005experimental}. However, an unexpected result is that for velocity ratios close to $1$, increasing $S_w$ reduces the range of parameter values for which stagnation occurs.

\begin{figure}
\centering
\begin{overpic}[width=0.35\textwidth]{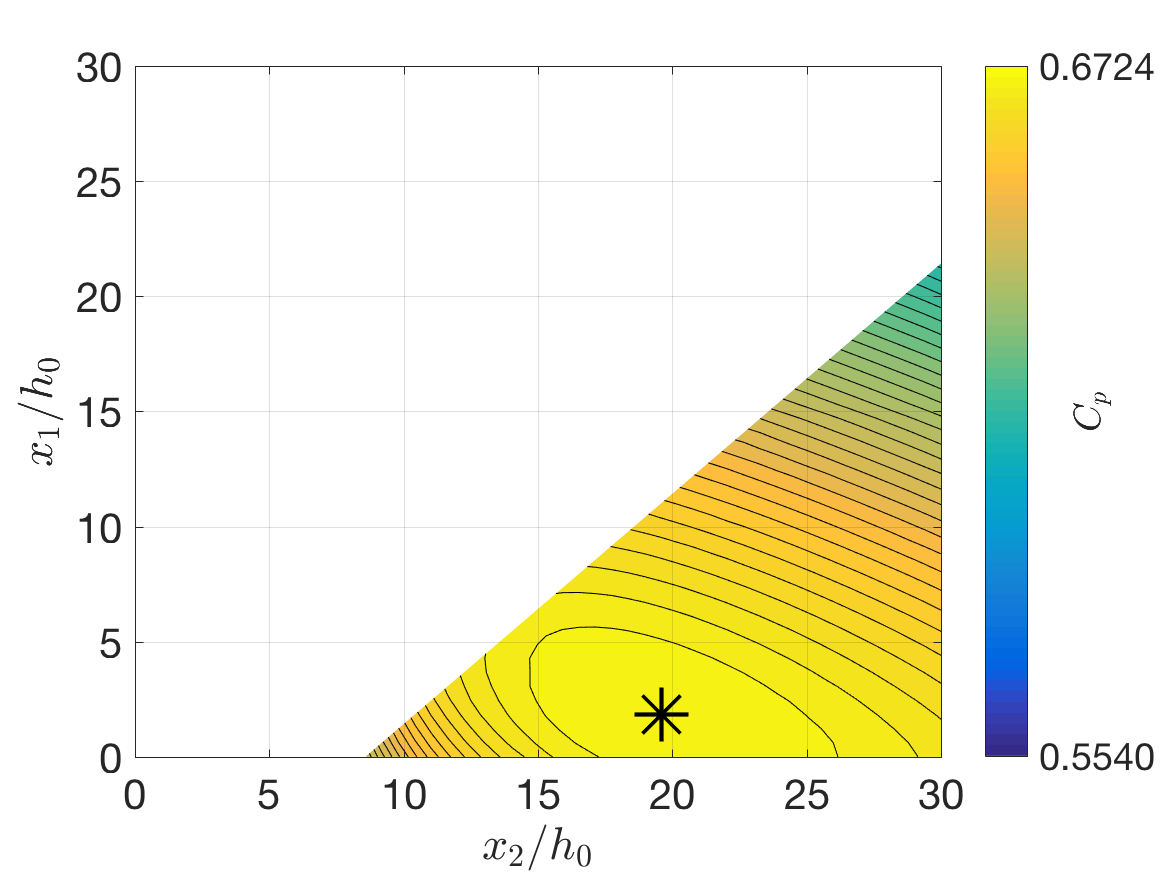}
\put(40,72){\large $S_w=0$}
\put (0,75) {(a)}
\put (12,50) {\includegraphics[width=0.23\textwidth]{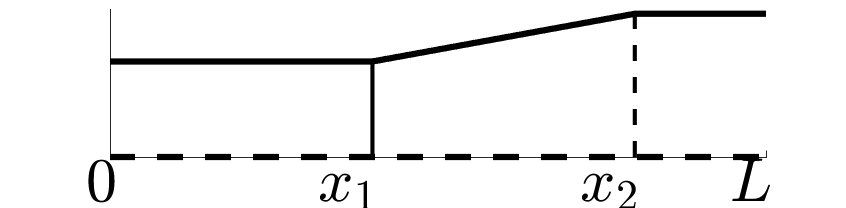}}
\end{overpic}
\begin{overpic}[width=0.35\textwidth]{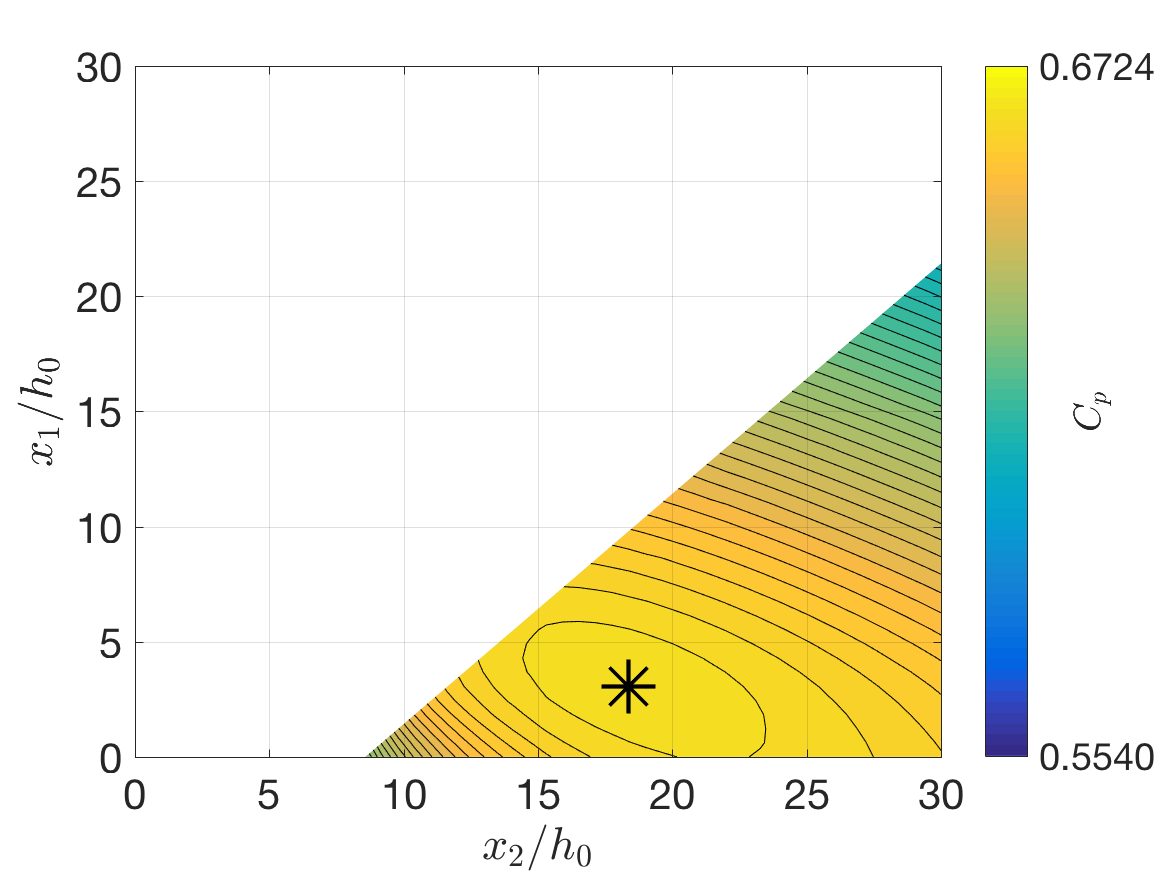}
\put(40,72){\large $S_w=0.27$}
\put (0,75) {(b)}
\end{overpic}
\begin{overpic}[width=0.25\textwidth]{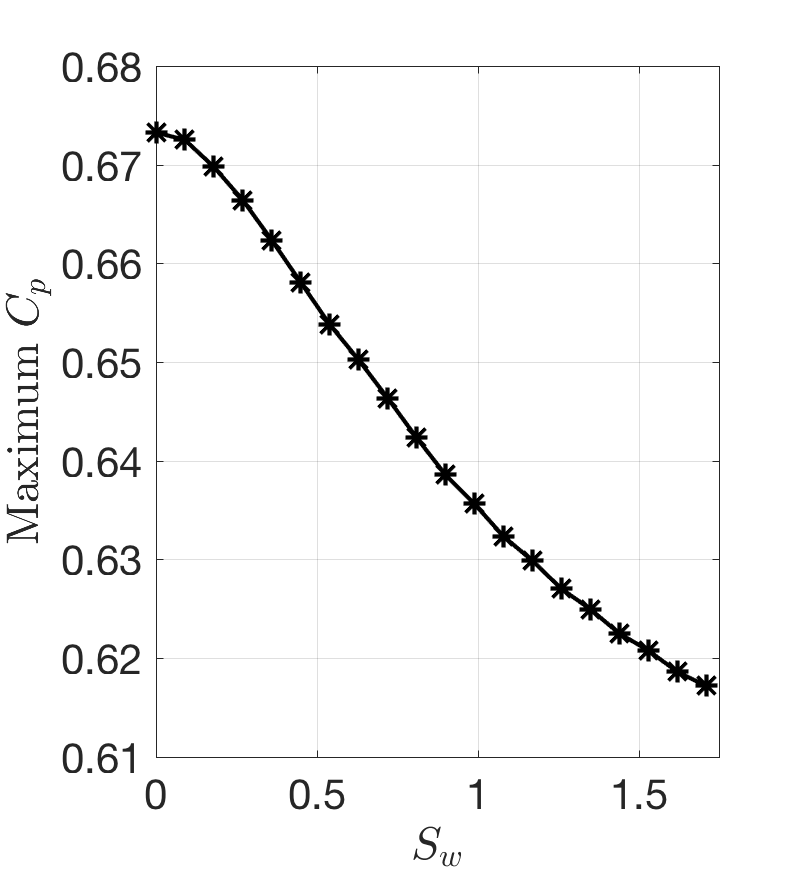}
\put(45,95){\large $C_p$}
\put (-5,95) {(c)}
\end{overpic}
\vspace{0.5cm}\\
\begin{overpic}[width=0.35\textwidth]{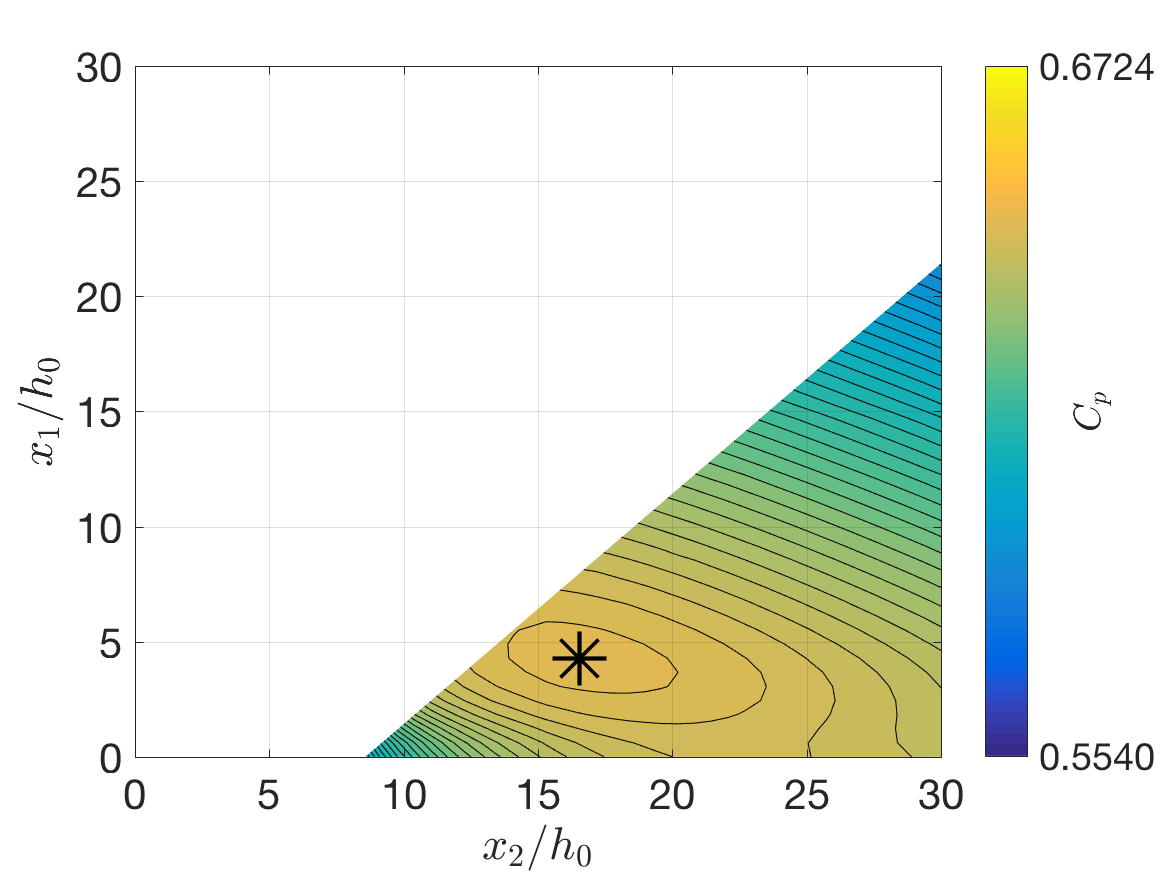}
\put(40,72){\large $S_w=0.67$}
\put (0,75) {(d)}
\end{overpic}
\begin{overpic}[width=0.35\textwidth]{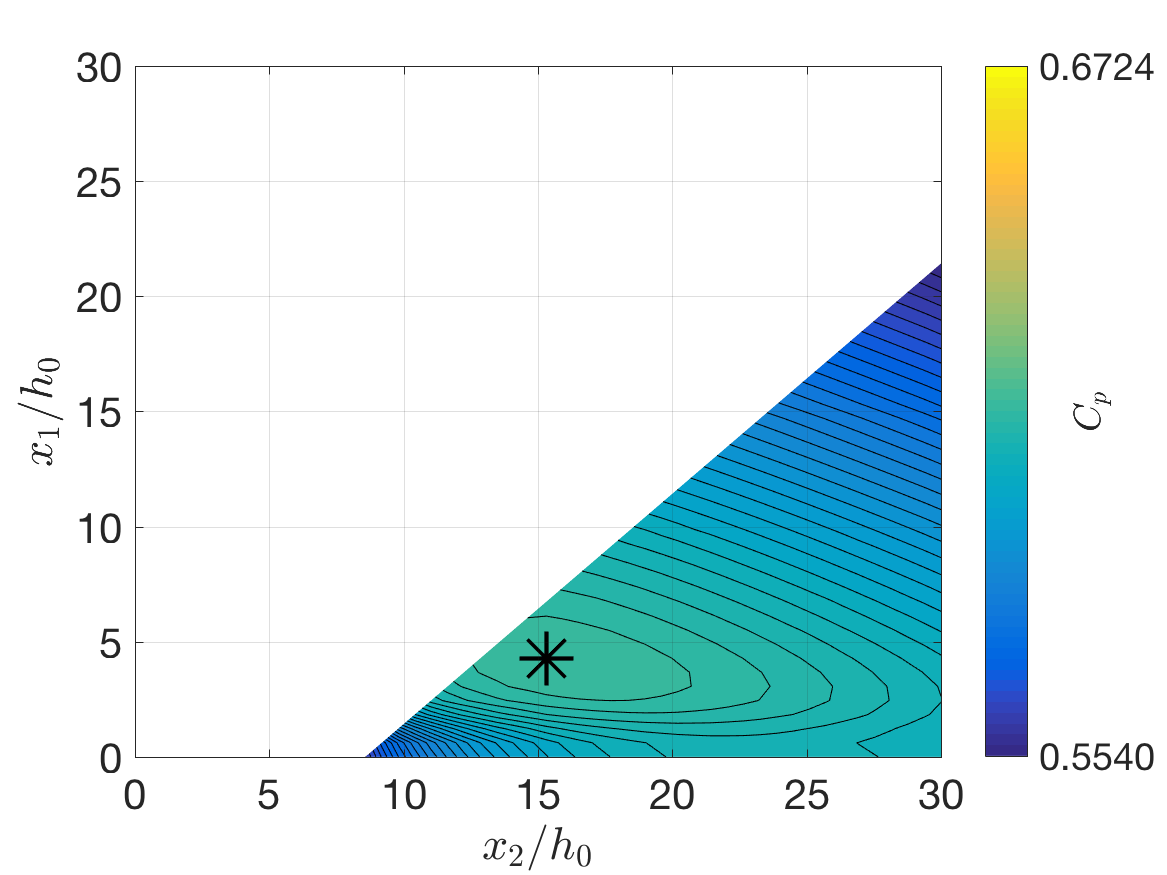}
\put(40,72){\large $S_w=1.71$}
\put (0,75) {(e)}
\end{overpic}
\begin{overpic}[width=0.25\textwidth]{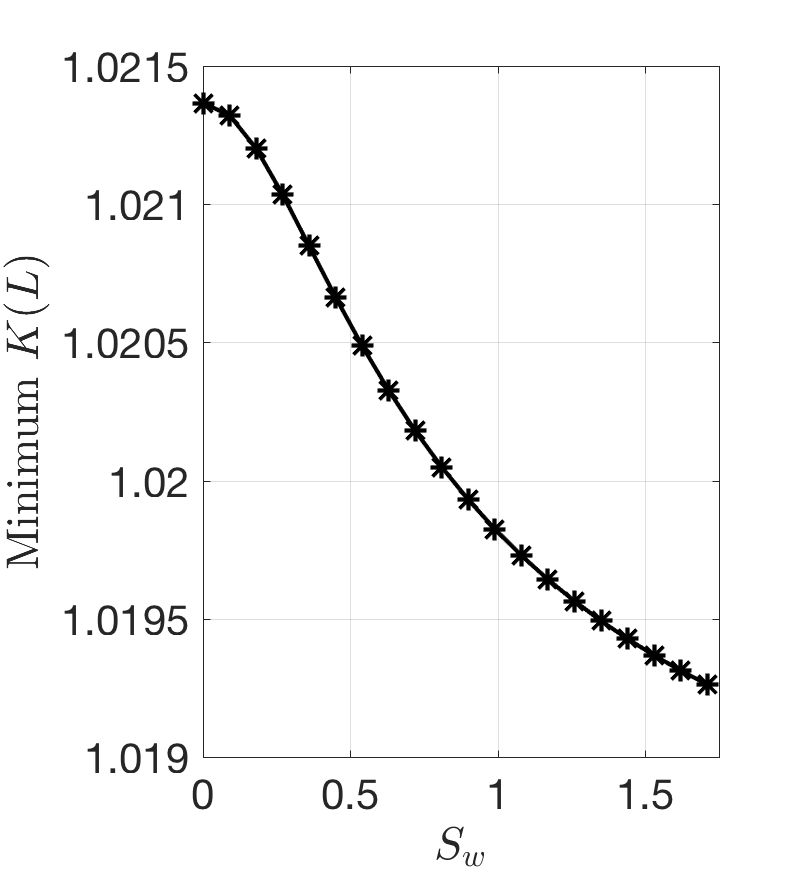}
\put(40,95){\large $K(L)$}
\put (-5,95) {(f)}
\end{overpic}
\caption{(a, b, d, e) Contour plots of pressure recovery coefficient $C_p$ for all permissible values of $x_1$ and $x_2$, a low-dimensional parameterisation of a diffuser shape which is divided into two straight sections separated by a widening section. Four different swirl number values are used and each contour plot has the same colour scale. Optimal shapes are indicated with a black asterisk. The optimum $C_p$ values for these and other swirl numbers are displayed in (c). A similar analysis is performed for the outlet kinetic energy flux profile factor $K(L)$ and the optimum values are displayed in (f). In all cases the inlet profile is given by $U_2(0)/U_0=0.5$, $h_1(0)=h_2(0)=h_0/2$.\label{contours}}
\end{figure}
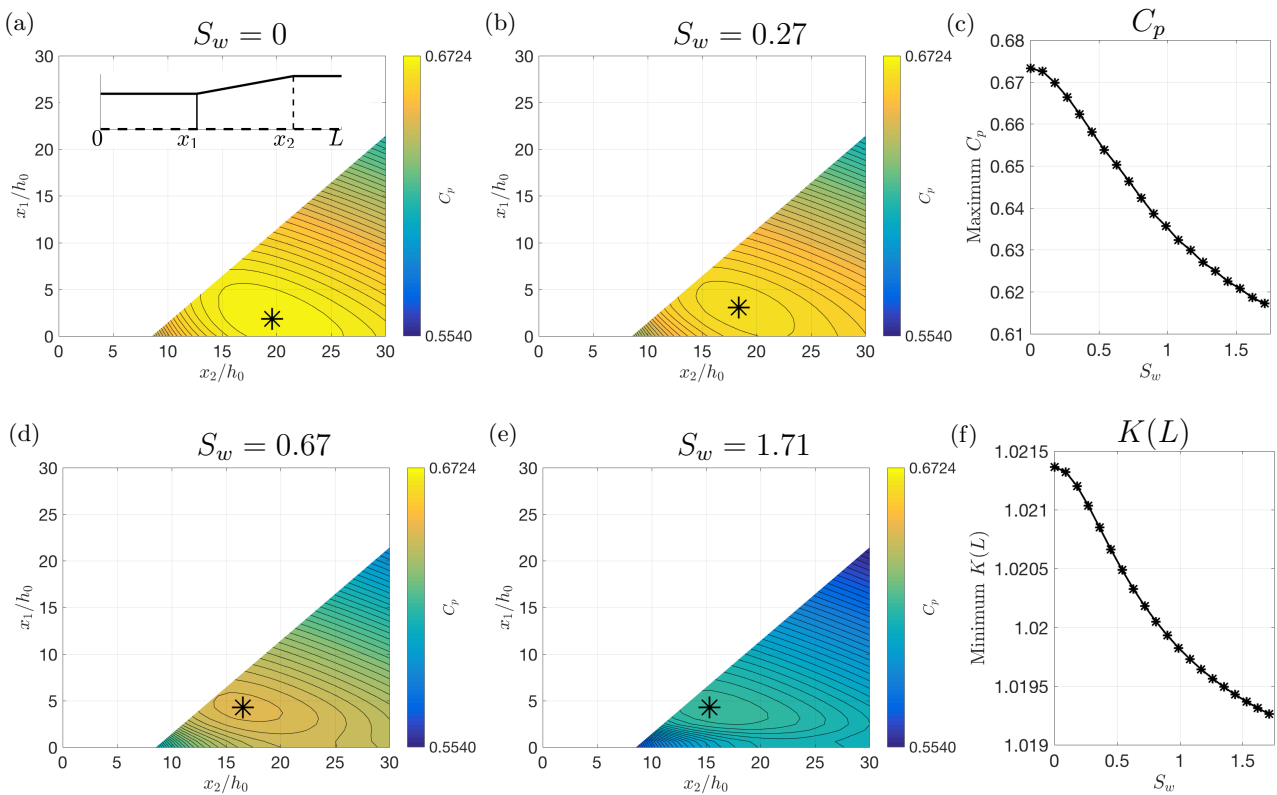

\subsection{Diffuser shape optimisation}

As a final investigation, we consider a shape optimisation problem in the presence of swirl. In order to illustrate the different effects of swirl, we consider optimising both $C_p$ and $K(L)$. \citet{benham2018optimal} used a similar approach to maximise pressure recovery in a flow diffuser (in the absence of swirl) by manipulating the pipe shape. In that study, a low-dimensional parameterisation of the shape is proposed, which consists of piecewise linear sections. Here, we make use of the same parameterisation, such that the pipe shape takes the form
\beq
h(x)=\begin{cases}
h_0:&0< x < x_1,\\
h_0 +\tan \alpha \lb x-x_1\rb :&x_1< x < x_2,\\
h_L:&x_2< x < L,\\
\end{cases}\label{shapepiece}
\eeq
where $x_1$, $x_2$ are the shape division points, $L$ is the diffuser length, $h_0$, $h_L$ are the inlet and outlet radii, and $\alpha = \arctan( (h_L-h_0)/(x_2-x_1))$ is the expansion angle (see Figure \ref{contours} (a)). We consider $h_0$, $h_L$, and $L$ fixed, whilst we consider $x_1$ and $x_2$ as decision variables.
For this class of shapes, we use our model to calculate both $C_p$ and $K(L)$ for all possible values of $x_1$ and $x_2$. 
The question of interest here is to what extent the addition of swirl to the inlet flow is beneficial for achieving mixing and pressure recovery within such a diffuser.
In each case, we choose a pipe which has non-dimensional length $L/h_0=30$ and expansion ratio $h_L/h_0=1.5$. The inflow is defined by $U_2(0)/U_0=0.5$ and $h_1(0)=h_2(0)=h_0/2$.
We exclude values of $x_1$ and $x_2$ which produce a pipe shape with an expansion angle larger than $3.5^\circ$ because it is expected that boundary layer separation can occur in such situations \cite{blevins1984applied}, which is not captured by our model. Hence, the plots are only shown in a triangular region. 

In Figure \ref{contours} we display contour plots of pressure recovery $C_p$ as a function of $x_1$ and $x_2$, for different values of the swirl number $S_w=0,\,0.27,\, 0.67,\,1.71$. For each value of $S_w$, we illustrate the optimum point with a black asterisk, and in Table \ref{Cptable} we display a table of the optimum points. 
{As swirl increases, the optimum point changes slightly, with $x_1$ increasing and $x_2$ decreasing. This indicates that diffusers with swirling core flows should be designed with a longer initial straight section and a shorter expanding section (with a wider expansion angle) than for non-swirling flows.}
However, the value of $C_p$ at the optimum point decreases with $S_w$, suggesting that a swirling core is always detrimental to pressure recovery, which is consistent with the results from Figure \ref{CpK} (a). 
We have also calculated $C_p$ for other values of $S_w$ and the maximum value indeed occurs at $S_w=0$ (see Figure \ref{contours} (c)). 
{These results can be explained by the fact that swirl creates a more non-uniform flow in the near-field, as seen in Figure \ref{CpK} (b). Therefore, swirling flows require a longer initial straight section to allow the flow to become more uniform before expanding, but the large drag due to this narrow section results in poor performance.}

Therefore, whilst swirl may improve pressure recovery when the flow is uniform, by reducing boundary layer separation \cite{fox1971effects, hah1983calculation}, it has the opposite effect on diffuser performance for non-uniform flow. 
It should be noted that our simple model does not account for boundary layer separation and, therefore, it cannot capture the positive effect of swirl on diffuser performance that is reported in the literature. However, since we restrict our attention to diffusers with small expansion angles, $\alpha <3.5^\circ$, there is little risk of boundary layer separation \cite{blevins1984applied}.

 \begin{table}
 \centering
 \begin{tabular}{|c||c|c|c||c|c|c|c}
\hline 
$S_w$ & $x_1$ & $x_2$ & $C_p$ & $x_1$ & $x_2$ & $K(L)$\\
\hline
0 & 1.8367  &  19.5918  &  0.6724&     0 & 8.5714 &   1.0214\\
0.27 &   3.0612 & 18.3673   & 0.6658 &    0 & 8.5714 &   1.0210\\
 0.67 &  4.2857 &  16.5306   & 0.6473 &   0 & 8.5714 & 1.0203\\
 1.71 &  4.2857 & 15.3061  & 0.6158 &    0 & 8.5714 &    1.0193\\
 \hline
 \end{tabular}
 \caption{List of optimum shape configurations for different swirl numbers. The left hand table corresponds to the maximisation of the pressure recovery $C_p$, whereas the right hand table corresponds to the minimisation of the outlet kinetic energy flux profile factor $K(L)$. \label{Cptable}}
 \end{table}

Similarly, we have calculated $K(L)$ as a function of $x_1$ and $x_2$ for different values of $S_w$. 
We do not show the corresponding contour plots here, but we display the minimum value of $K(L)$ for each value of $S_w$ in Figure \ref{contours} (f), and a table of the optimum points in Table \ref{Cptable}.
{We find that the optimum shape is independent of swirl, having no initial straight section ($x_1=0$) and a very short widening section (with a large expansion angle).}
However, as $S_w$ increases the minimum value of $K(L)$ decreases, indicating that swirl enables the optimal shape to produce a more uniform outflow, which is consistent with the results from Figure \ref{CpK} (b).
We can conclude from these results that, whilst swirl makes the outflow more uniform, it produces greater pressure losses in doing so.

\section{Summary and concluding remarks}\label{secsumm}

We have developed a simple model for confined co-axial flow with a core swirling region, which shows good agreement with CFD, whilst being computationally cheap. 
The model depends on two parameters, the friction factor $f$, which is well approximated with the Blasius relationship \cite{blasius1913ahnlichkeitsgesetz,mckeon2005new}, and the spreading parameter $S_c$, which is fitted to CFD calculations, giving a value {$S_c=0.13$} which is within the range of reported values \cite{benham2018turbulent}. 

The model is useful for predicting essential aspects of the flow behaviour, such as the shear layer growth and pressure recovery. We use this model to characterise the criteria for which a recirculation region appears along the pipe axis, and at what distance downstream. We show that, whilst adding swirl to the core region is a good mechanism for increasing shear layer growth rates, it also increases pressure losses. Therefore, we conclude that a swirling core is advantageous for situations where mixing the flow over a short length scale is important, such as in combustion chambers, whereas it is disadvantageous for situations where pressure recovery is important, such as in flow diffusers.
{The low computational cost of the model makes it ideal for design purposes, and we illustrate this by using the model to optimise the shape of a diffuser for different values of the swirl number.}

We show that an extended version of the model, which includes a boundary layer at the pipe wall and a symmetry layer at the pipe axis, shows even better agreement with CFD calculations than the simple model. Whilst unnecessary for the conclusions made using the simple model, the extended model provides more detailed and realistic flow predictions.

For future work, the effect of different swirl distributions for the core flow, such as a $q$ vortex \cite{gilchrist2005experimental}, could be investigated. Furthermore, this work could be extended to situations in which the outer annular flow is also rotating, not necessarily in the same direction as the core flow.\\


This publication is based on work supported by the EPSRC Centre for Doctoral Training in Industrially Focused Mathematical Modelling (EP/L015803/1) in collaboration with VerdErg Renewable Energy Limited and inspired by their novel Venturi-Enhanced Turbine Technology for low-head hydropower.\\

\bibliographystyle{apsrev4-1}
\bibliography{bibfile}

\end{document}